\documentclass[aps,prb,singlecolumn,superscriptaddress, longbibliography, nobibnotes]{revtex4-2}
\usepackage{graphicx}
\usepackage{color}
\usepackage{float}
\usepackage{amsmath}
\usepackage{todonotes}

\usepackage[breaklinks]{hyperref}
\hypersetup{
	pdfnewwindow=true,      % links in new window
    colorlinks=true,       % false: boxed links; true: colored links
    linkcolor=blue,          % color of internal links
    citecolor=blue,        % color of links to bibliography
    filecolor=blue,      % color of file links
    urlcolor=blue           % color of external links
}

\begin{document}
\title{Cathodoluminescence, light injection and EELS in STEM: From comparative to coincidence experiments.}
\author{Luiz H. G. Tizei}
\email{luiz.tizei@cnrs.fr}
\affiliation{Univ. Paris-Saclay, CNRS, Laboratoire de Physique des Solides, 91405, Orsay, France}

\author{Yves Auad}
\affiliation{Univ. Paris-Saclay, CNRS, Laboratoire de Physique des Solides, 91405, Orsay, France}

\author{Florian Castioni}
\affiliation{Univ. Paris-Saclay, CNRS, Laboratoire de Physique des Solides, 91405, Orsay, France}

\author{Mathieu Kociak}
\affiliation{Univ. Paris-Saclay, CNRS, Laboratoire de Physique des Solides, 91405, Orsay, France}

%Acronyms to define: MOF, EELS, PL CL, EDS, HAADF, FIB, LN2, SI.
%Acrynums defined: ROI: region of interest

\begin{abstract}
Electron spectroscopy implemented in electron microscopes provides high spatial resolution, down to the atomic scale, of the chemical, electronic, vibrational and optical properties of materials. In this review, we will describe how temporal coincidence experiments in the nanosecond to femtosecond range between different electron spectroscopies involving photons, inelastic electrons and secondary electrons can provide information bits not accessible to independent spectroscopies. In particular, we will focus on nano-optics applications. The instrumental modifications necessary for these experiments are discussed, as well as the perspectives for these coincidence techniques.
\end{abstract}

\maketitle

\section{Introduction}
\label{sec::intro}
Electron spectroscopies have been used for decades to explore the chemistry, electronic structure, vibrational properties and optical response of materials. A non-exhaustive list includes: electron energy-loss spectroscopy (EELS), energy-dispersed X-ray spectroscopy (EDS), cathodoluminescence (CL), Auger spectroscopy, secondary electron (SE) spectroscopy and electron energy-gain spectroscopy (EEGS). For spatially resolved experiments, these techniques are implemented in electron microscopes (EM), which allow for the generation of electron probes down to the sub-atomic scale. Depending on the experiment to be performed, these techniques can be implemented on scanning electron microscopes (SEM) or scanning transmission electron microscopes (STEM). EELS can also be implemented with high spatial resolution in transmission electron microscopes (TEM), a technique named energy-filtered TEM (EFTEM). These spectroscopies have applications in many fields, ranging from chemistry and material science to optics. This review is focused on electron spectroscopies implemented in SEM and STEM, and specifically on their uses in temporal synchronization in the context of nano-optics experiments. 

\begin{figure}
\centering{\includegraphics[width=0.6\textwidth]{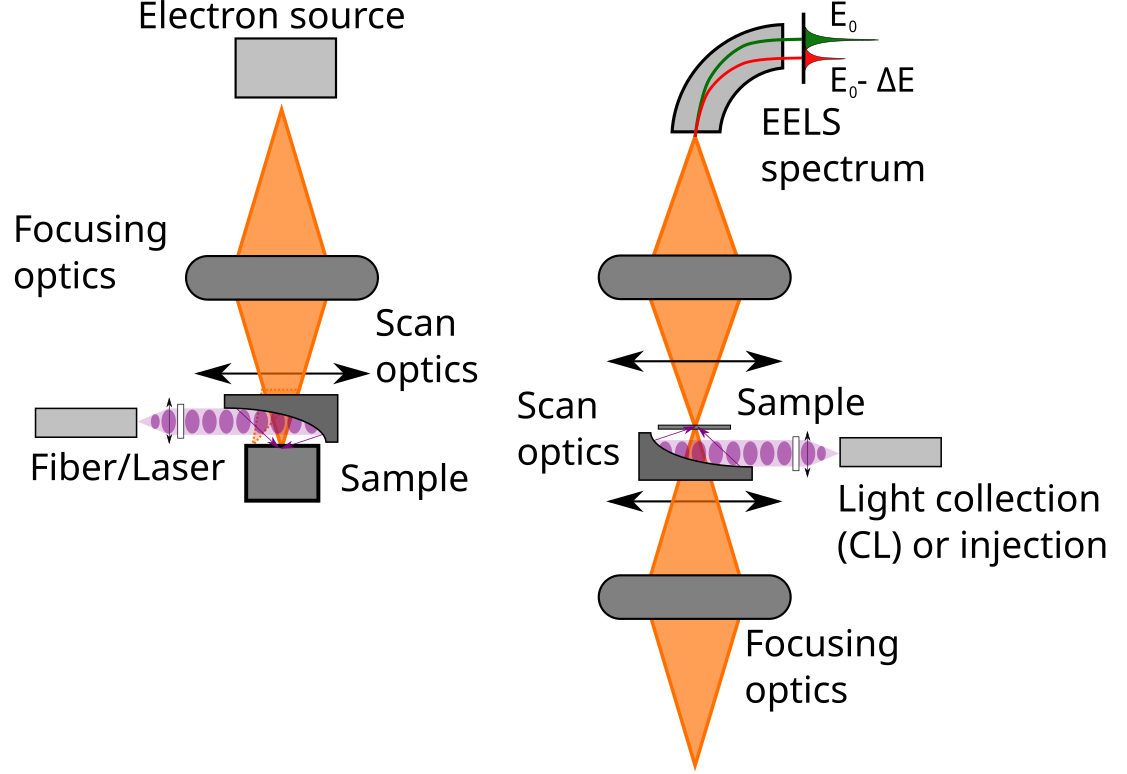}}
\caption{\textbf{Sketch of SEM/STEM for spectroscopies linked to inelastic electron light scattering.} The fundamental parts of the system are an electron source, focusing, scanning and projection optics, a light collection/injection system and an electron spectrometer. Nano-optics synchronized experiments can be performed in both STEM or SEM microscopes.
\label{fig::Stem_CL}}
\end{figure}

In SEM and STEM (Fig. \ref{fig::Stem_CL}) a focused electron probe is scanned on a sample, and different spectroscopy information bits are acquired. Some of these will be described in Section \ref{sec::comparative}. In the past, SEM and STEM were distinguished by their electron beam kinetic energy: below 30 keV for SEMs and above 300 keV for STEMs. As technology advanced, especially aberration correctors at low voltages, and thin sample preparation improved, STEM experiments also at low voltages became a reality \cite{Suenaga2010}, blurring the lines between both technologies.

In short, SEM/STEM are composed of an electron source, an electron optical system to generate a focused electron probe that can be scanned on the sample, and an ensemble of supplementary electron optics and detector to perform the different spectroscopies. Details of the required electron optics can be found in the literature \cite{Hawkes2008, Kohl2008}.

The objective of this review is to describe temporal coincidence or synchronized experiments joining different electron spectroscopies or laser irradiation and how these can provide information not accessible to each of the electron spectroscopies independently. First, we will describe the independent spectroscopies in Section \ref{sec::comparative}. In Section \ref{sec::coincidence_setup}, possible sets of modifications of SEM/STEM to implement synchronized measurements for nano-optics are discussed, followed by implementations of these experiments in Section \ref{sec::synchronized}. Specifically, we will discuss photon(CL)-photon(CL) in Section \ref{sec::synchronized::photon-photon}, electron(EELS)-photon(CL) in Section \ref{sec::synchronized::photon-electron}, photon(injection)-electron(EELS) in Sections \ref{sec::synchronized::injection-eels} and \ref{sec::synchronized::nanothermometry} and electron(EELS)-electron(Auger or secondary) in Section \ref{sec::synchronized::electron-electron}. 

\section{Comparative electron spectroscopies for nano-optics}
\label{sec::comparative}

The field of nano-optics begins when the size and geometry of nano-objects, such as nanoparticles or heterostructures, strongly influence their optical properties. This arises from the strong confinement of two main types of excitations within materials. Broadly speaking, the first category includes excitations whose dynamics are primarily governed by the Schr\"odinger equation. This is the case for electron-hole pairs, excitons, or bound states associated with point defects in semiconductors or insulators. In this context, quantum confinement, which is a phenomenon typically occurring over the scale of a Bohr radius (from a few angstroms to several tens of nanometres), is one of the dominant effect relevant to nano-optics.

The second category involves so-called photonic excitations, which are essentially described by Maxwell’s equations. These include modes in insulators (guided modes, whispering gallery modes, photonic crystal modes, etc.) on one hand, and plasmonic excitations (localized plasmons, plasmonic crystal modes, etc.) on the other. One can also add to those the Cherenkov and transition radiations, the latter having a high degree of importance in the next sections of this article. Here, the confinement relevant to nano-optics is classical, and the typical confinement scales are set by the wavelength of the light in the materials - on the order of a few hundred of nanometres for the photonic systems, and several tens of nanometres for surface plasmons in the visible range.

The interaction between a fast electron and matter is commonly described in simplified terms as either incoherent or coherent \cite{Abajo2010,Kociak2014}. In the former case, the electron interacts with the bulk of the sample, predominantly creating a bulk plasmon at relatively high energy (typically tens of eV, well above the optical range). This excitation quickly relaxes into electron-hole pairs, which then diffuse and thermalize \cite{Meuret2015}. In semiconductors and insulators, these pairs may subsequently form mixed electron-hole excitations with energies in the optical range. In the coherent case, the electromagnetic field generated by the electron excites the system’s modes in phase.

The two historical electron-based spectroscopies, EELS and cathodoluminescence (CL), can be used to probe the optical properties of nanomaterials and trace them back to the underlying excitations. An EELS spectrum reflects all coherent excitations generated in the material - excitons, bulk plasmons, or surface plasmons alike. Optically, EELS corresponds to extinction spectroscopy, encompassing both absorption and scattering phenomena. In the pedagogical case of a dipolar excitation with polarizability $\alpha(\omega)$ at energy $\hbar\omega$, the EELS signal in an aloof configuration is proportional to $Im(\alpha(\omega))$ , just like the optical extinction cross-section \cite{Kociak2014, Losquin2015}. It is worth noting that the polarizability is the observable of interest in optical experiments on nanoparticles explaining the value of EELS when probing optical properties of nanoparticles. It turns out that on bulk material, EELS permits to measure the full dielectric constant of bulk material on very large energy ranges, an historical application of this technique that won’t be discussed \cite{Abajo2010}.

In contrast, a cathodoluminescence spectrum reflects different physics and excitations depending on the context. When the interaction is coherent, as in the case of photonic excitations discussed further below, and the excitation is radiative (e.g., a dipolar surface plasmon), some light is emitted in the far field. In this regime, CL is analogous to optical scattering, with spectra proportional to $|\alpha(\omega)|^2$  for dipolar excitations. When the electron interaction is incoherent, the light emission process resembles non-resonant photoluminescence \cite{Mahfoud2013, Kociak2017}.

One key advantage of using fast electrons over photons is the ability to form highly focused electron beams on very small areas. It is clear that performing EELS and CL experiments on the same object, or even within the same object, provides access to fundamentally different physical information bit which are worth comparing.

Finally, a third form of spectroscopy offers spectral information similar to CL. When an optical system is irradiated by an intense laser beam, a stimulated electron energy gain and loss results, giving rise to sidebands at integer multiples of the laser energy across the spectrum. This is the so-called photon induced near field electron microscopy (PINEM) effect. In the linear regime, a PINEM spectrum reduces to a single gain and loss peak whose intensity is proportional to $|\alpha(\omega_L)|^2$, where $\omega_L$ is the laser frequency. Measuring this intensity as a function of the laser energy provides a spectroscopy known as EEGS, which probes scattering physics, much like CL for plasmons \cite{Asenjo2013}. EEGS and PINEM, which are technically closely related to coincidence-based methods discussed later in this article, will be addressed in greater detail subsequently.

\begin{figure} [H]
\centering{\includegraphics[width=0.4\textwidth]{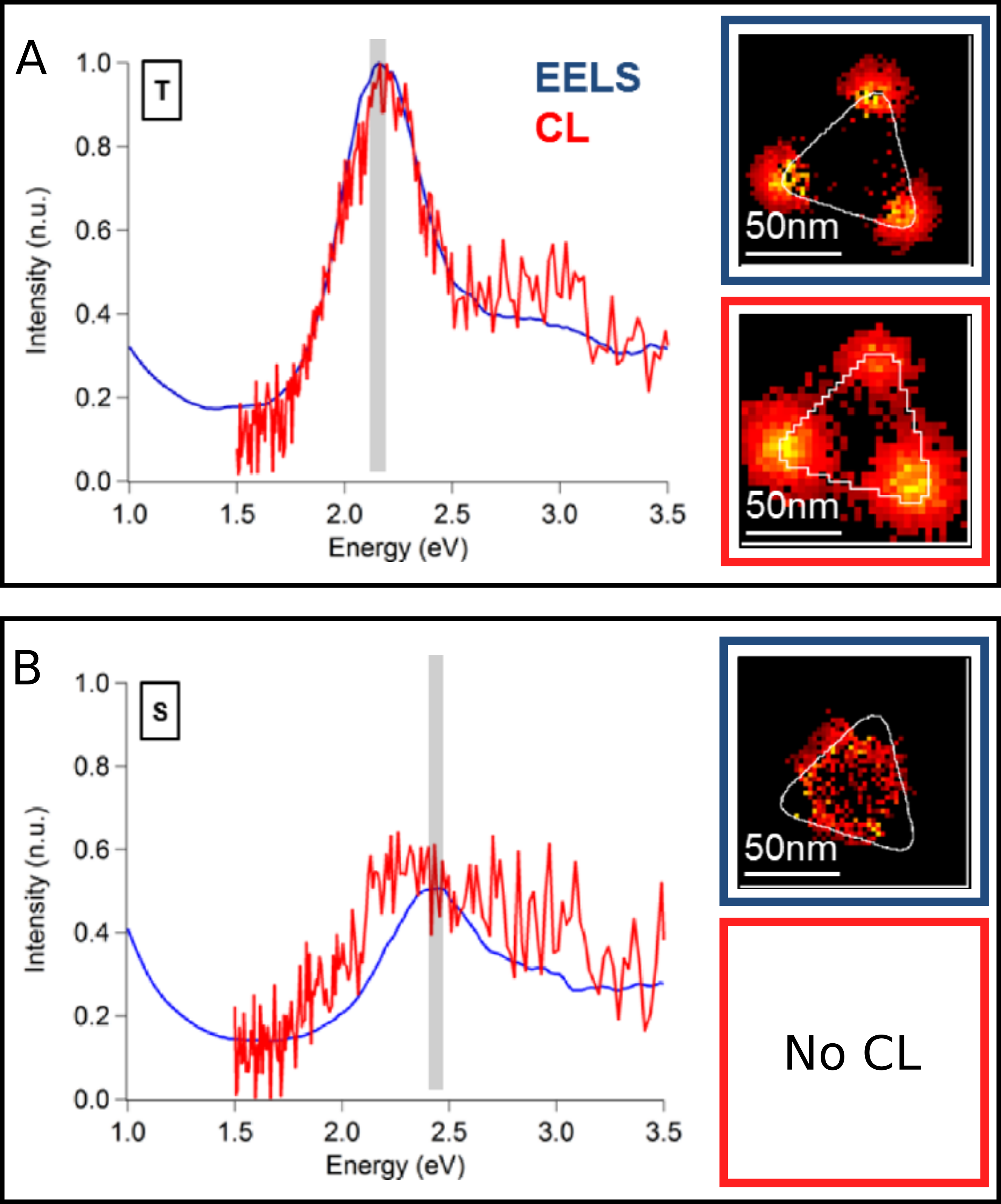}}
\caption{\textbf{Correlative EELS/CL experiments on plasmonic gold nanotriangles.} (a) Spectra and fitted maps for the dipolar, tip (T) mode. (b) Same for the hexapolar, side (S) mode. Note the absence of fittable S peak in CL. Adapted from \cite{Losquin2015}.
\label{fig::CL-EELS_Plasmons}}
\end{figure}

In this section, we provide an overview of experiments comparing EELS, CL, and, where applicable, PINEM or EEGS on the same objects. Fig. \ref{fig::CL-EELS_Plasmons} reflects these considerations in the case of surface plasmons in plasmonic nanoparticles \cite{Losquin2015}. EELS and CL spectra measured at the tip of a triangular gold nanoparticle (Fig. \ref{fig::CL-EELS_Plasmons}A, left) both show a dominant peak corresponding to the dipolar mode of the nanostructure. This becomes even clearer when plotting the amplitude of Gaussian-fitted peaks in the spectra versus position (Fig. \ref{fig::CL-EELS_Plasmons}A, right), which shows maxima at the tips. This is characteristic of dipolar modes in triangular systems \cite{Nelayah2007}. The spatial variations of the dipolar mode are very similar between the two spectroscopies. This is expected since the EELS spatial variation closely resembles that of the z-component of the local density of electromagnetic states - total for EELS and radiative for CL \cite{Abajo2008, Losquin2015b}. The peak positions in the spectra are also identical. However, this is not a general result: for dissipative excitations, the absorption and scattering cross-sections differ ($Im(\alpha(\omega)\neq|\alpha(\omega)|^2)$), and their coincidence here is accidental (see full discussion in Ref. \cite{Losquin2015, Kawasaki2016}).

Differences between EELS and CL become more apparent for higher-order modes, such as the hexapolar mode (Fig. \ref{fig::CL-EELS_Plasmons}B, left), which is detected in EELS with a clear spatial signature (Fig. \ref{fig::CL-EELS_Plasmons}B, right), but not in CL. This is because in small particles like these (less than 60 nm typical in edge length), only dipolar modes are radiative \cite{Novotny2012}. However, due to retardation effects non-dipolar modes can acquire radiative character in larger particles \cite{Schmidt2017}. In general, as nanoparticle size increases, the CL-to-EELS cross-section ratio increases for higher-order modes. This has been quantified in the case of plasmonic disks \cite{Schmidt2017}.

This behaviour arises because the EELS cross-section scales with the object volume (for a sphere, polarizability $\alpha(\omega)$ and therefore $Im(\alpha(\omega))$, scales as the cube of the radius), whereas the CL cross-section scales with the square of the volume, as expected from the $|\alpha(\omega)|²$  dependence. Instrumentally, this makes EELS more suitable for small particles, while CL becomes advantageous for larger ones (see \cite{Kociak2014} for a full discussion). The comparison between EELS and CL is thus highly fruitful for identifying whether an excitation is radiative or non-radiative (see for example the case of supposed to be dark modes \cite{Schmidt2017,Guo2019}).

Historically, this comparison was challenging, as it required both high spectral resolution EELS and a high numerical aperture CL detector within the same transmission electron microscope (TEM). Consequently, such comparisons were long limited to studies of plasmons or transition radiation, whose spectral width is large enough so that they do not require a monochromator to be studied. They also often involved different microscopes. Using different microscopes introduces complications: differing acceleration voltages (especially between STEM for EELS/PINEM and SEM for CL \cite{Liebtrau2021}), vacuum levels, and potential sample alterations \cite{Husnik2013} (aging, etc.) as the samples are moved from one EM to another, all of which may hinder accurate comparisons of spectral features across techniques. Moreover, absolute calibration - necessary for comparing EELS, PINEM/EEGS, and CL energies - is a challenge whether within the same microscope or not \cite{Losquin2015}.

\begin{figure}
\centering{\includegraphics[width=0.4\textwidth]{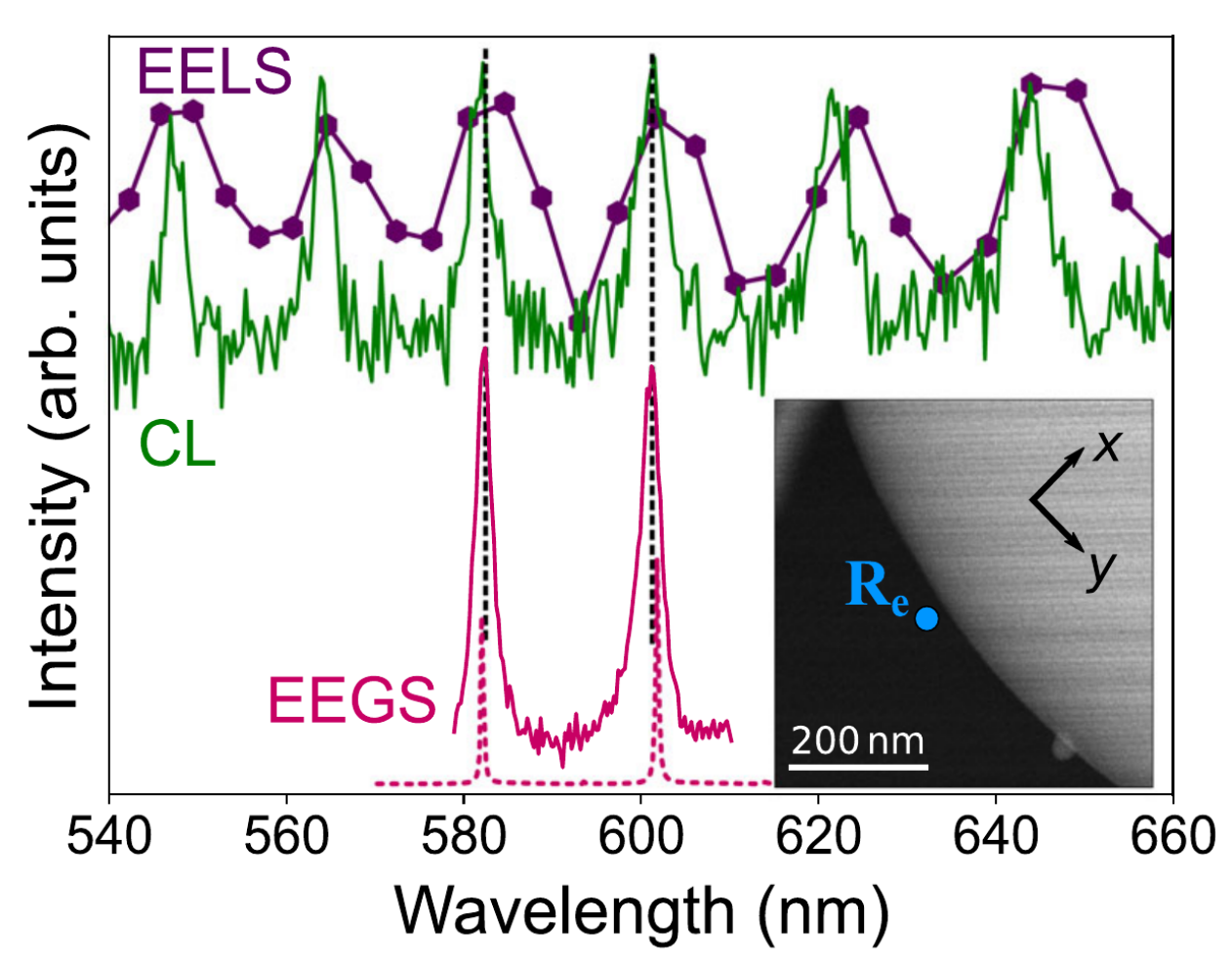}}
\caption{\textbf{Comparison between EEGS, CL, EELS for whispering gallery modes in a silica sphere of diameter 4 µm.}  Data have been acquired at the same point close to the sphere surface (marked $R_e$ in the ADF image in the inset). Adapted with permission from \cite{Auad2023}.
\label{fig::EEGS}}
\end{figure}

The advent of monochromated and highly monochromated microscopes has enabled access to energy ranges relevant to photonic systems \cite{Cha2010, Thomas2013, Bezard2024, Auad2021}. It has thus been demonstrated that, in such systems where dissipation is minimal, EELS, CL, and EEGS all exhibit resonances at exactly the same energy, within experimental error \cite{Auad2021,Auad2023}. This is shown in Fig. \ref{fig::EEGS} for a whispering gallery mode excitation. However, each technique has a distinct application scope: EELS, even with expensive monochromators, remains the least spectrally resolved and is not polarization-sensitive in normal scattering configuration, but it can probe photonic excitations from the far-infrared to the UV \cite{Auad2021}. Polarization effects can be detected in EELS by changing the scattering geometry, for example, selecting specific scattering vectors to form the EELS spectrum \cite{Krehl2018} or by placing samples at specific angles for polarized measurements \cite{Schattschneider2012}. Also, the use of phase shaped electron beams allows for polarization dependent measurements \cite{Lourencco2021}, what has been confirmed experimentally for surface plasmons \cite{Guzzinati2017}. Compared to EELS, CL is inexpensive and easy to implement. But it has low sensitivity, especially for weakly radiative excitations\cite{Auad2023}, and a limited range of detection. EEGS is much more complex to implement, is restricted to small energy ranges, but offers excellent signal-to-noise and very high spectral resolution \cite{Auad2023, Henke2021}.

\begin{figure}
\centering{\includegraphics[width=1\textwidth]{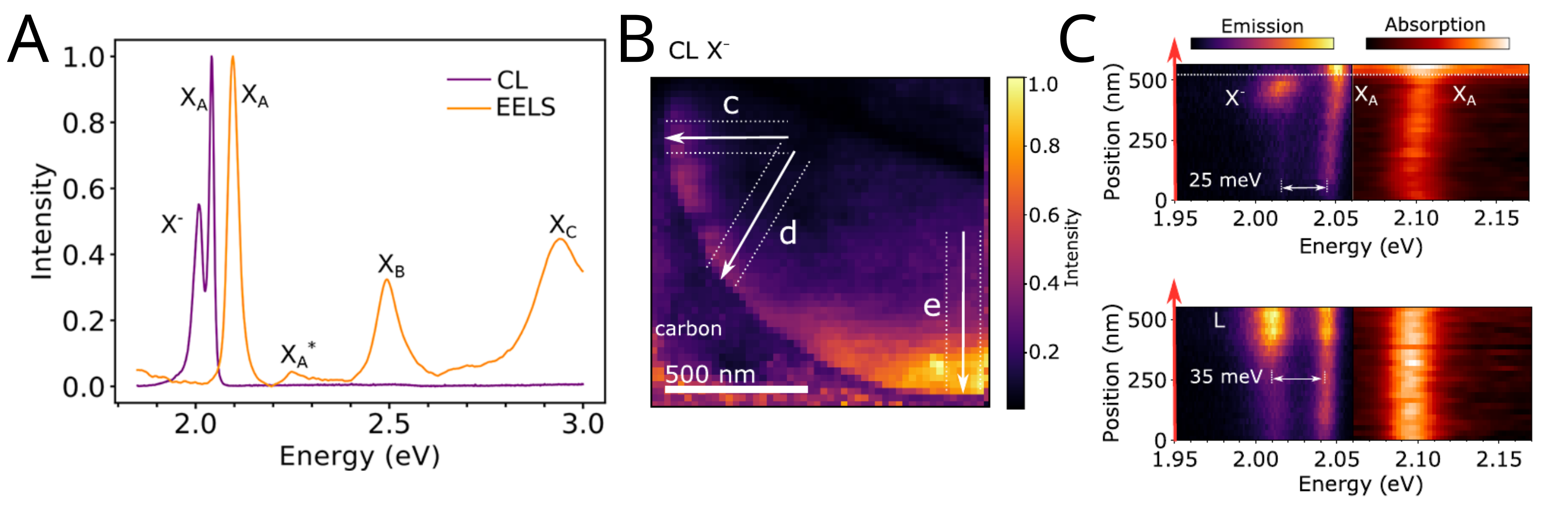}}
\caption{\textbf{Comparison between CL and EELS for a WS$_2$  monolayer encapsulated in hBN}  (a) An EELS and CL spectrum of a hBN/Wse2/hBN heterostructure. In the EELS spectrum, the exciton $X_A$, $X_B$ and $X_C$ and the excited exciton $X_A^*$ are seen. In the CL spectrum the neutral exciton $X_A$ and the charged exciton (trion) $X^-$ emission are observed. (b) Filtered CL map at the trion energy. (c) line profiles along the arrows marked “c” (upper) and “e” (lower). Adapted with permission from \cite{Bonnet2021}.
\label{fig::EELS-CL_exciton}}
\end{figure}

Combining monochromators with CL systems in a STEM has also enabled comparison of absorption and luminescence properties of electron-hole-type excitations since only the early 2020s \cite{Bonnet2021,Woo2024}. Fig. \ref{fig::EELS-CL_exciton} illustrates the value of such comparisons. Here, EELS and CL spectra are acquired and compared for a monolayer of WS$_2$ encapsulated in hBN \cite{Bonnet2021}. This sample preparation not only generates enough electron-hole pairs for a decent CL signal \cite{Zheng2017} but also enhances the optical properties of the TMD monolayers in general \cite{Shao2022}. In the absence of significant scattering, the extinction cross-section essentially reduces to absorption. Indeed, EELS reveals absorption by highlighting the main excitons ($X_A$, $X_B$, $X_C$) in TMDs (Transition Metal Dichalcogenides) which occur over a large energy range \cite{Wang2018}. Conversely, CL probes luminescence properties, here dominated by the $X_A$ exciton and a lower-energy trion $X^-$. Comparing CL and EELS spectra thus provides complementary physical insights. Fig. \ref{fig::EELS-CL_exciton}B shows the trion intensity mapping, and Fig. \ref{fig::EELS-CL_exciton}C illustrates spatial variations in absorption and emission spectra at the nanoscale. The constant Stokes shift for exciton A indicates that the trions’ spectral variations arise from emission, not absorption.

It is worth noting that no direct comparison with optical absorption and emission experiments has been made on the same semi-conductor sample; however, qualitative agreement has been found with similar samples \cite{Woo2024}. The understanding of differing Stokes shifts between electron- and photon-based experiments remains an open question.

As with photonic excitations, such EELS/CL comparisons remain rare due to the stringent experimental requirements. Notable examples include EELS/CL comparisons in quantum-confined structures, where confinement signatures have been observed only in CL \cite{Bachu2024}, as the absorption oscillator strength of the quantum confined structure is below the current detection limits, and differences between EELS and CL in studying exciton-guided mode coupling in thick WSe$_2$ flakes \cite{Taleb2022}. Similar comparisons would be highly valuable for technologically relevant materials like hybrid perovskites\cite{Hou2021}, for example in understanding the mechanisms underlying the Stokes shift to optimize device performance while decoupling the absorption and emission bands.

\section{Experimental setup for nano-optics coincidence electron spectroscopy }
\label{sec::coincidence_setup}
Time-averaged electron spectroscopies have contributed to solve many problems in materials’ physics. Yet, many aspects remain inaccessible to these techniques, especially those linked to excitations and their dynamics. For example, linewidths in time-averaged spectroscopies can be affected by excitations lifetime, but a complete understanding of an excitation dynamics can only be observed with time-resolved techniques. Moreover, the link between absorption and emission energy is usually lost in electron spectroscopies, due to the broadband nature of electron excitation \cite{Abajo2010}.

\begin{figure} [H]
\centering{\includegraphics[width=0.7\textwidth]{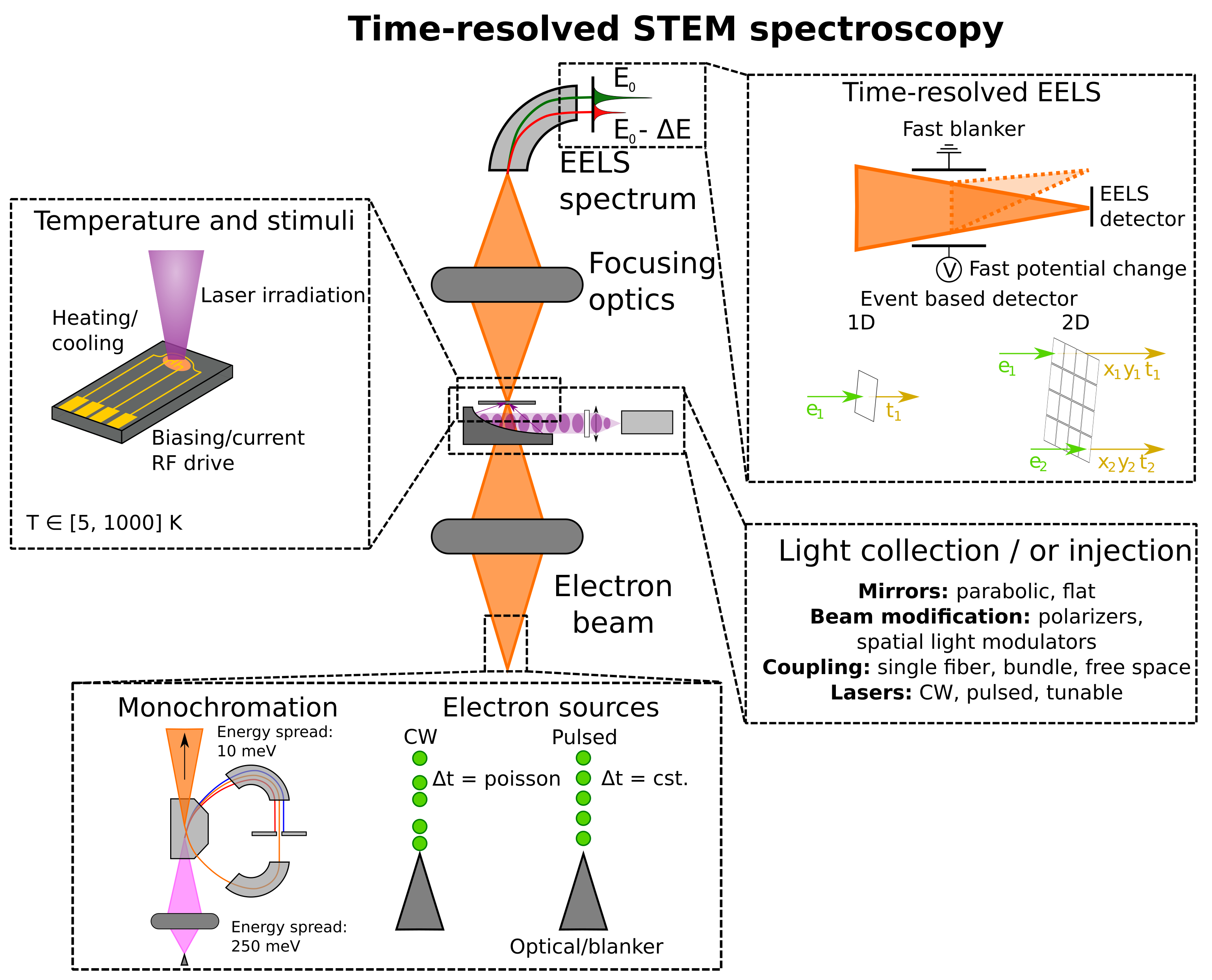}}
\caption{\textbf{Experimental setup for time-resolved and synchronized experiments in electron microscopes and details about the necessary instrumentation for uses with a continuous or a pulsed electron source.} 
\label{fig::TimeResolvedNanooptics}}
\end{figure}

These limitations can be tackled by time-resolved spectroscopies, which can be achieved by different synchronization methods. Synchronized electron spectroscopy experiments have been explored in the past in order to extract information unavailable in time-averaged measurements \cite{Kruit1984, Ahn1984, Haak1984, Pijper199}. Some of these have attracted recently renewed interest, especially because of the development of modern 2D array event-based electron detectors, focusing on nano-optics \cite{Feist2022, Varkentina2022} or material science \cite{Jannis2021} topics. In Fig. \ref{fig::TimeResolvedNanooptics}, we sketch some of the different options impacting coincidence experiments for nano-optics, which can be adapted depending on the targeted applications.

\subsection{Electron sources}
\label{sec::coincidence_setup::sources}

First and foremost, one must make choices concerning the electron beam generation (Fig. \ref{fig::TimeResolvedNanooptics}, lower left). Different electron sources exist \cite{Williams2009}, but the brightest ones are cold field emitters (FEGs), which allow for the generation of sub-angstrom probes with the highest energy resolution: 250 meV without monochromation. To further improve the energy resolution, electron monochromators have been developed \cite{Krivanek2014, Boerrnert2023} with the current record spectral resolution at around 3 meV for 20 keV electrons \cite{Dellby2023}. Pulsed sources have been developed using electromagnetic blankers, radio frequency cavities or photoemission from metallic tips, achieving temporal resolution down to the sub-fs scale \cite{Arbouet2018}. Continuous (with Poisson statistics) electron sources have the benefits of keeping the EM optics unchanged, ensuring optimal performance and high current. 

Of primary interest for this review, temporal resolution is still possible to access for these sources through time-resolved detection, either using fast blankers after the sample and/or time-resolved detectors. The post-specimen fast-blanker technology has been used to perform time-resolved electron spectroscopy with 5 ns temporal resolution \cite{Das2019}. This pump-probe approach, achieved by the synchronization of the electron arrival time on a detector with other signals, has been used to perform ns-resolved electron spectroscopy with 20 meV spectral resolution at 200 keV \cite{Auad2023} and 100 keV \cite{Castioni2025}.

Both schemes, pulsed sources or time-resolved detection, require synchronization electronics. Typically, this involves the synchronization of different clocks or the distribution of a common clock among different electronics. The second solution is preferable, as it can ensure better timing stability over longer temporal ranges. Synchronization of different events are needed: photon-photon (for second order correlation measurements\cite{Brown1956}), electron-photon (for EELS-CL  \cite{Varkentina2022}, EELS-EDS \cite{Jannis2021} or EELS-injection \cite{Henke2021, Auad2023}) or electron-photon-photon (for energy resolved second order correlation measurements).

A key difficulty in time-resolved experiments is the amount of signal theoretically available in a small time bin. As an example, for a 1 pA current poissonian electron beam, which is usually referred to as “continuous” and typical for experiments with currently available event-based detectors, there are on average only $\approx$0.006 electrons in a 1 ns time bin. Realistically, experiments with this temporal resolution can therefore only be performed stroboscopically, and temporal transformations or modifications of the sample need to be cyclic.

\subsection{Detectors}
\label{sec::coincidence_setup::detectors}
Concerning detectors, different solutions are possible (Fig. \ref{fig::TimeResolvedNanooptics}, right). For electrons,  scintillators coupled to photomultiplier tubes (PMTs) were a solution in the past, which was limited by their temporal response (10s ps – 1 ns, depending on the technology used) \cite{Kruit1984, Ahn1984}. They were used essentially as single pixel detectors, which were great for imaging (high-angle annular dark-field, HADF, bright field, BF or SE, etc.) or serial EELS, but not for parallel EELS. Currently, event-based detectors\cite{Auad2022, Auad2024} and microchannel plates (MCP) coupled to delay line detectors that both reach ~1 ns time resolution \cite{Jannis2019} are parallel detectors that can be adapted on EELS spectrometers. For photons, a larger number of solutions exists, ranging from PMTs, avalanche photo-diodes and superconducting nanowire detectors, which reach 10s ps temporal resolution. 

For X-rays, this is possible with silicon drift detectors (SDD) detectors, despite their current poor temporal resolution \cite{Jannis2019, Jannis2021}. For Auger and SE, PMTs coupled to scintillators in electron spectrometers have been used. Finally, UV-vis-IR photons can be collected using mirrors or fibers and detected using time-resolved photon detectors. For maximal collection efficiency without compromising spectral resolution elliptical and parabolic reflectors have been used in TEMs \cite{Pennycook1980, Yamamoto1984}. To avoid timing errors due to materials and mode dispersion in optical fibres, free space propagation of photons to the detectors should be preferred. Typically, these effects lead to timing errors of the order of 100 ps. 

\subsection{Photons detection/collection and injection}
\label{sec::coincidence_setup::photon_collection}

For light injection/collection reflector systems (flat, parabolic) can be used (Fig. \ref{fig::TimeResolvedNanooptics}, center right). These have been used extensively in the TEM community for PINEM and EEGS experiments \cite{Barwick2009, Feist2015, Das2019}. Parabolic geometry can ensure a higher fluence (Total energy per surface area, J/m$^2$) than their flat counterparts. In these mirrors light polarization control is less straightforward than for flat mirrors. Also, the presence of a hole and their off-axis geometry does not allow for the generation of a gaussian spot. Still, focusing down to a diffraction limited $\approx$µm spot is feasible \cite{Auad2023}. For easy polarization and laser spot geometry, flat mirrors should be preferred, at the cost of a larger illumination spot size. 

In injection experiment in the ns temporal range, optical fibres can be used for ease of alignment. To maximize the fluence, monomode fibres are the best option. For example, in PINEM/EEGS experiments, the laser fluence is the important quantity in maximizing signal. If multimode fibres are used, the total energy deposited will most probably be higher, but not the PINEM signal, leading to sample heating or damage. Another variable in injection experiments is the pulse width of the laser source. Shorter laser pulses allow for higher fluence for a given average power. Therefore, PINEM can be achieved more easily with a fs-laser source in comparison with a continuous one. In choosing a laser pulse temporal width ($\Delta t$), one should remember that it is ultimately linked to the pulse energy width ($\Delta E$) by $\Delta t \Delta E > 0.3 eV.fs$  \cite{Abajo2025}. The minimum energy width of a 100 fs pulse is around 3 meV, that of 1 ps pulse 0.3 meV and that of 1 ns pulse 0.0003 meV. As a final consideration for laser sources, tunability can be achieved in different ways: solid state diodes for continuous sources, dye-lasers in the nanosecond range, non-linear fibre-based white sources in the ps range and finally optical parametric amplifiers in the sub 100 fs range.

Still concerning light collection and injection, modifications of the light polarization, phase and shape can be achieved by polarizers and amplitude and phase modulators. For example, spatial light modulators have been used to modify a laser profile in PINEM experiments \cite{Konevcna2020}, which in the future could be used for electron phase modification for geometric aberration correction \cite{Konevcna2020, Nekula2025}.

\subsection{In situ experiments}
\label{sec::coincidence_setup::insitu}

Laser irradiation can also be used to modify the sample. Nanoparticles can be deformed and grown by laser beams\cite{Werner2008}. Similar ideas can be used to study photocatalysis, possibly coupling imaging, diffraction and spectroscopic measurements. Laser-irradiation synchronized to EELS detection is a possible way to study samples out-of-equilibrium (for example, the dynamics of excitons in semiconductors). Laser irradiation can increase the temperature of a sample, as observed by the modification of core-hole excitations in carbon nanotubes \cite{Rossouw2013}. This process can be used to clean samples, including hBN and graphene monolayers \cite{Tripathi2017}. Finally, time-resolved irradiation experiments in the nanosecond range have shown that the temperature dynamics of different materials can be tracked using EELS spectroscopic signatures in metal and semiconductors \cite{Castioni2025}. Modification of a sample temperature using a focused laser has the added benefit of creating a localized temperature gradient without the need of complicated sample preparation compared to Joule heaters described later, and with a small total power. This minimizes sample drift and the amount of time needed for thermalization before spatially resolved measurements.

Despite disadvantages, Joules heaters built into the sample or its support (Fig. \ref{fig::TimeResolvedNanooptics} left) are an effective way to induce temperature differences \cite{Mecklenburg2015, Lagos2018, Idrobo2018, Yang2025}. In addition to heating, electrical contacts can be used to induce changes in samples using voltages, which can be significantly faster than those induced by heating and, hence, followed using stroboscopic methods. The magnetization can also be modified using specialized holders and followed in time if cyclic.
In addition, electrical contacts can be used to measure the current induced on a sample, for example a p-n junction, due to the impinging electron beam (Electron beam induced current, EBIC). The synchronization methods described here can, in principle, allow for synchronized experiments to measure the energy-loss-dependence of the generated current (EELS-EBIC experiments) similarly to what has already been done for EDS-EELS \cite{Jannis2021} and CL-EELS \cite{Varkentina2022}.
For in situ experiments, sample preparation is a key aspect, as accessing the sample with contacts, light input and other means of stimuli is needed. Sample preparation for EM is already an art in itself, where thin (10-200 nm typically) samples are required. Many of the sample modification methods, especially those based on current and voltages benefit from the optical and electronic lithography and focused ion beam (FIB) techniques. We refer the interest reader to specialized reviews and texts \cite{Ayache2010}.

Heat, voltage and current supplying devices with micrometres in size are produced on dedicate micro-electromechanical systems (MEMS). These have to be connected to the sample holder, through which they are connected to the outside of the EM. A versatile solution to this problem is the use of insulating cartridges to which MEMS can be fixed to and connected using wire-bonding.

\section{Synchronized experiments in CL, EELS and EEGS}
\label{sec::synchronized}

\begin{figure} [H]
\centering{\includegraphics[width=0.7\textwidth]{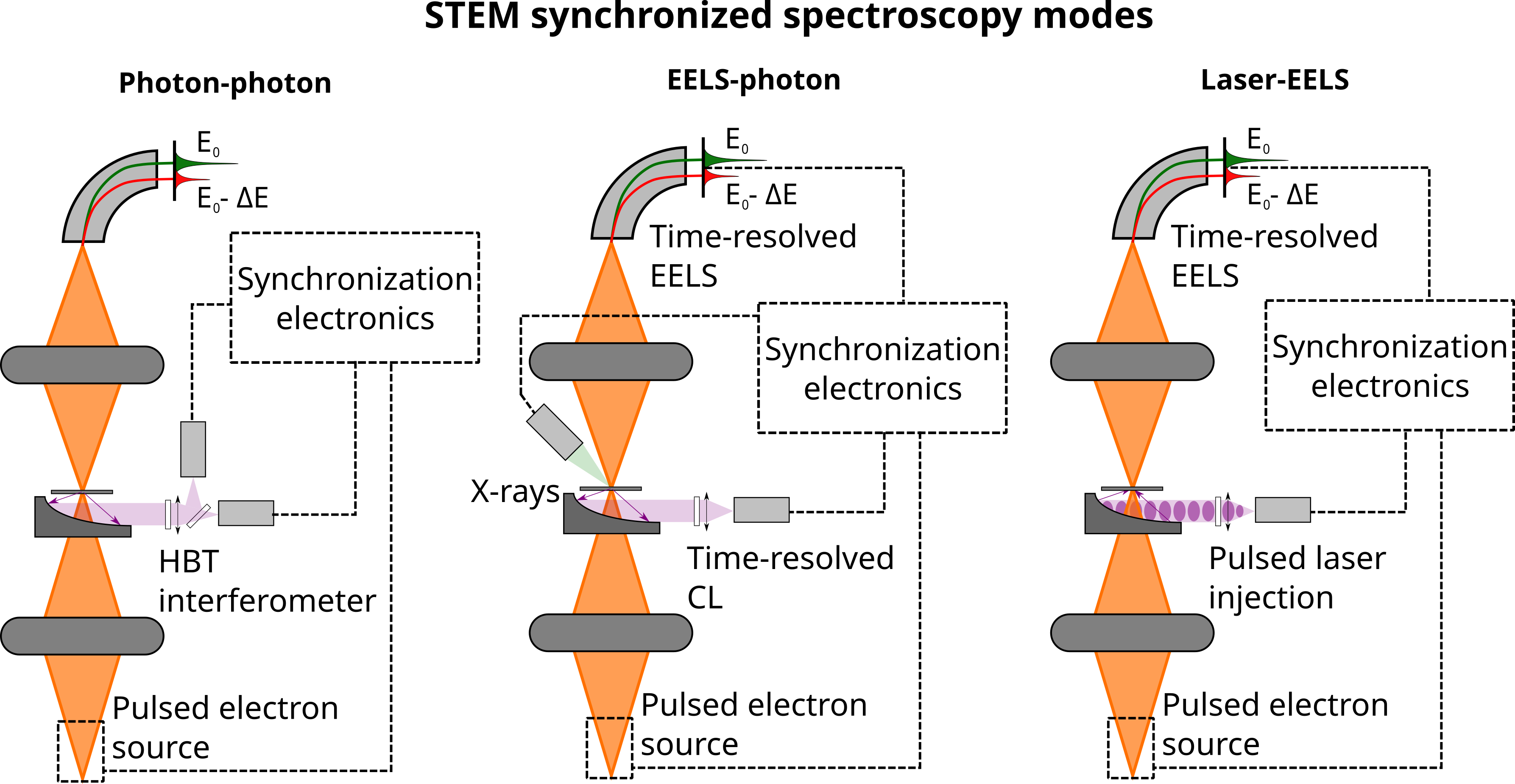}}
\caption{\textbf{Setups for temporally synchronized spectroscopies:} (a) CL-CL; (b) EELS-EDS, EELS-CL; (c) Laser-EELS for PINEM/EEGS or nanothermometry. All the techniques can be implemented in microscopes with time-resolved electron sources for improved temporal resolution. Timing information from different detectors, including the emission time of electrons in pulsed electron sources, are time-tagged in a common clock using correlation electronics.
\label{fig::coincidencesketch}}
\end{figure}

Synchronized experiments between different signals can explore quantities not accessible to time-averaged experiments, as for example, the temporal statistics of the light emitted by a given defect. In the next subsections, experiments concerning photon-photon (Fig. \ref{fig::coincidencesketch}A) electron-photon collection (Fig. \ref{fig::coincidencesketch}B) and photon injection-electron (Fig. \ref{fig::coincidencesketch}C) experiments will be described.

\subsection{Photon-photon synchronization}
\label{sec::synchronized::photon-photon}

\begin{figure} [H]
\centering{\includegraphics[width=0.7\textwidth]{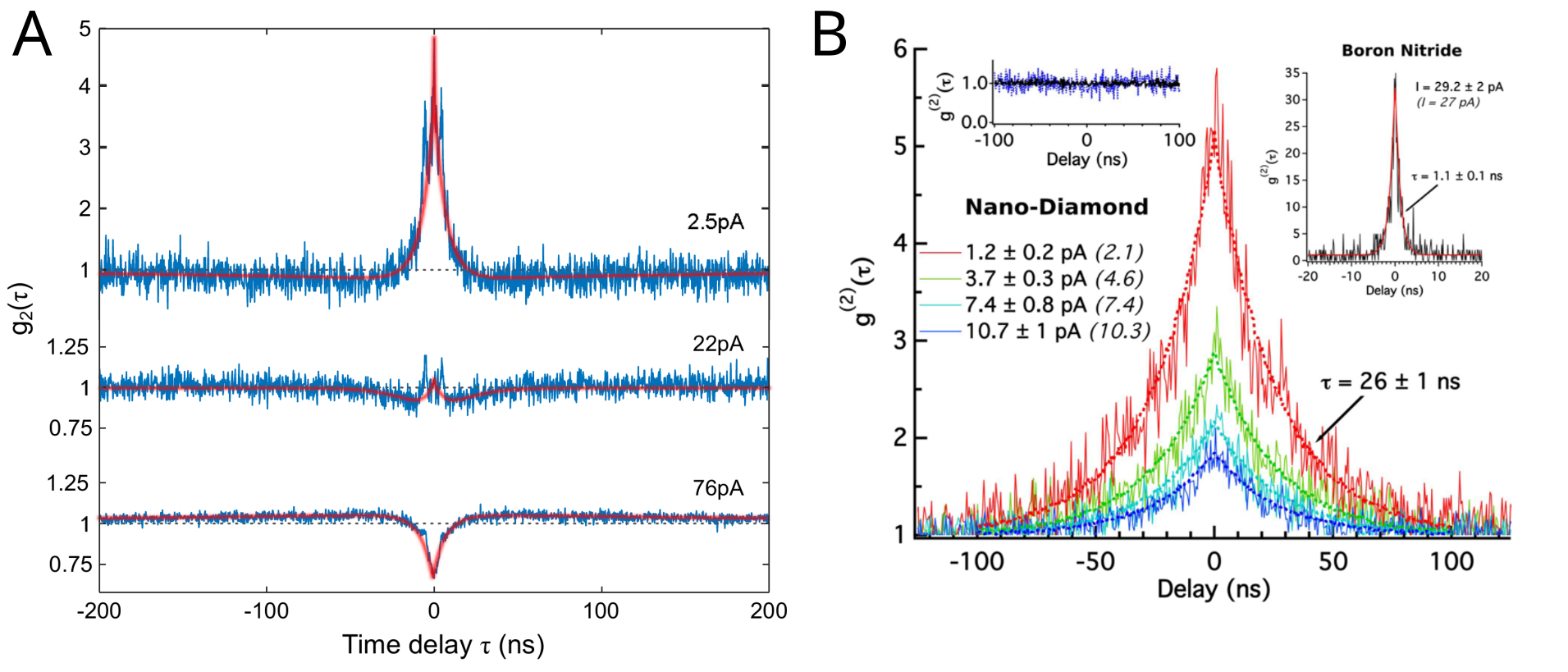}}
\caption{\textbf{Antibunching and bunching in CL experiments:} (a) Transition from light antibunching (high current, lower curve) to light bunching (low current, upper curve) observed in CL experiments on Ge centres in diamond. Adapted with permission from \cite{Fiedler2023}. (b) Light bunching as a function of electron beam current for a diamond nanoparticle containing a large number of NV centres in CL experiments. The upper right inset shows light bunching in CL for an hBN flake containing 4.1 eV defects. The upper left inset shows a  for a diamond nanoparticle from light excitation (PL). Measured lifetimes values are given. Adapted with permission from \cite{Meuret2015}.
\label{fig::antibunching}}
\end{figure}

The second order correlation function, $g^{(2)} (\tau) = \langle I(\tau)I(t+\tau)\rangle/ \langle I(\tau)\rangle \langle I(t+\tau)\rangle$, encodes information about intensity ($I$) correlations at different time delays ($\tau$) in a light beam. In quantum optics, it is used to distinguish the states of light. The value of this function at zero-time delay distinguishes classical ($g^{(2)}(0) = 2$), coherent ($g^{(2)}(0) = 1$) and single photon ($g^{(2)}(0) = 0$) light source (Fig. \ref{fig::antibunching}). Typically, this function is measured using a Hanbury Brown and Twiss (HBT) light intensity interferometer \cite{Fox2010}, a device composed of at least a beam splitter, two time-resolved photon detectors and a correlator, which measures the time-delay between photon detection events \cite{Brown1956}.

In EM, this type of interferometer was used for the first time to detect single NV$^0$ centres in diamond nanoparticles \cite{Tizei2013}, a well-known single photon source \cite{Gruber1997}. Since then, HBT interferometers in EMs have been used to identify a new single photon source in hBN emitting at 4.1 eV \cite{Bourrellier2016} and to observe single photons emitted from Ge-related defects in diamond \cite{Fiedler2023} (Fig. \ref{fig::antibunching}A).

In addition to light antibunching, HBT experiments in EMs show bunching (Fig. \ref{fig::antibunching}) \cite{Fiedler2023,Meuret2015, Feldman2018}. This effect was first observed using beta decay as the electron source \cite{Borst1999}, interpreted as an effect of the creation of multiple electron-hole pairs per incident primary electron and proposed as an effective way to measure lifetimes. Later, when observed in electron microscopes, the possibility of measuring lifetimes at high spatial resolution was demonstrated \cite{Meuret2016}. This technique has been explored to measure the lifetime, excitation probability and quantum efficiency of emitters in different experimental setups \cite{Finot2021}. As the bunching signature increases at lower excitation currents, a transition between light bunching and antibunching has been observed for Ge defects in diamond \cite{Fiedler2023}, as predicted previously \cite{Meuret2015}.  statistical model has been developed to describe the impact on the measured second order correlation function of the statistics of each process in the chain leading to photon emission \cite{Yanagimoto2025}.

\subsection{Photon-electron synchronization}
\label{sec::synchronized::photon-electron}
Electron-photon coincidence measurements can be achieved through synchronized EELS-CL (near IR-UV photons) or EELS-EDS (X-ray photons) experiments. These can provide details about the excitation, relaxation and de-excitation processes which are hidden in time-averaged comparative experiments. 

Photon-electron coincidence measurements have been already demonstrated in the 1980s-1990s, both for visible \cite{Ahn1984} and X-ray photons \cite{Kruit1984}. In the visible range, in addition to the measurement of excited state lifetimes, this method allowed to determine the linear dependence of the number of CL photons emitted as a function of the energy loss \cite{Ahn1984} in cerium doped yttrium silicate (phosphor P47). An increase in photon production near the band edge of the phosphor as well as a reduction at the bulk plasmon energy were reported \cite{Ahn1984}. For X-ray photons, EELS background subtraction was the initial target application, which was limited by the collection efficiencies for X-rays and electrons \cite{Kruit1984}. Despite promising outlooks, these techniques were not explored further due to technological limitations.

The development of event-based hybrid electron detectors \cite{Jannis2021, Auad2022, Auad2024}, and specifically the introduction of the Timepix3 detector for electron detection, motivated a renewed interest in photon-electron coincidence experiments. This new class of detectors allowed for the parallel detection of the time delay between electrons and photons with temporal resolution down to 1.3 ns and over a wide range of energy losses, significantly speeding up the data acquisition in comparison to experiments on filtered energy windows. This, summed with the improved quantum efficiency of hybrid electron detectors and those of photon detectors, made experiments which were unfeasible with the former technologies eventually accessible. Here, we note that detecting 30-300 keV kinetic energy electrons with high signal to noise ratio is easier than detecting visible or UV range photons. However, only recently has the broader electron microscopy and spectroscopy community become interested in and gained access to high SNR detectors, such as hybrid direct electron detectors.

\begin{figure} [H]
\centering{\includegraphics[width=0.9\textwidth]{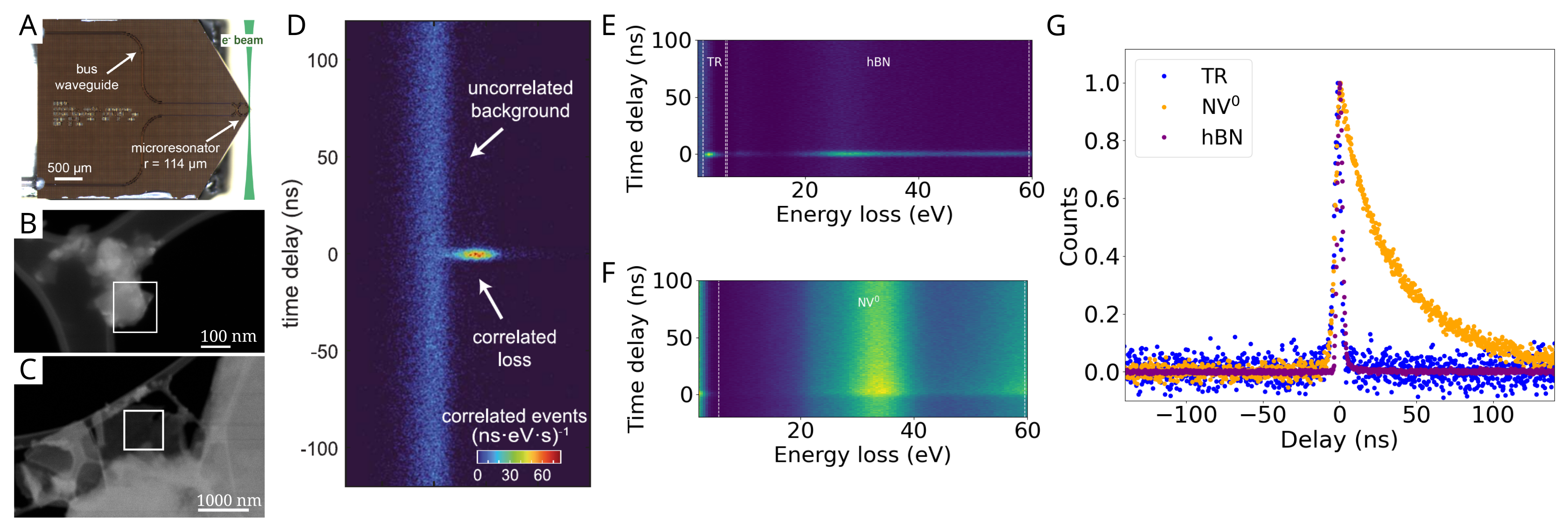}}
\caption{\textbf{EELS-CL synchronized experiments:} (a) An image of optical ring resonator structure and its light incoupling guides. Electrons passing next to this structure (green sketch in (a)) have a probability of generating photons in the resonator. Histogram of the electron-photon time-delay and energy lost by the electron (d) shows a large number of coincidences for zero-time delay and energy matching the modes in the structure. Adapted with permission from ref. \cite{Feist2022}. (b-c) ADF images of nanodiamonds and an hBN flake. EELS-CL synchronized measurements were used to generate the time-delay/energy-loss 2D histograms in (e) and (f). The coincidences for large time delays in nanodiamonds match the known lifetime of NVs in diamond ($>$20 ns). This can be seen in details in (g), where the decay profiles extracted from (e) and (f) for transition radiation and the defects in BN and diamond are shown. Adapted with permission from \cite{Varkentina2023}.
\label{fig::CLE}}
\end{figure}

For X-rays, Jannis et al. \cite{Jannis2021} demonstrated that synchronized measurements improved the model-free background subtraction both for EELS and EDS, including for peaks superimposed in energy. This methodology becomes interesting for trace elements detections, for which small signals are expected in comparison to those from a matrix. The main limitation for this technique is the large timing uncertainty (400 ns) in X-ray detection time using SDD, which arise due to the detectors large surface area \cite{Jannis2021}. 

For visible photons, synchronized EELS-CL experiments have been used to demonstrate the generation of electron-photon pairs following the inelastic scattering of an electron on micrometre scale optical cavity \cite{Feist2022} (Fig. \ref{fig::CLE} A and D). Correlated electron energy-loss and photon emission are only observed on a specific mode of the optical cavity (at 1.55 µm or 0.8 eV).

A similar setup, using a Timepix3 detector and PMTs for synchronized EELS-CL experiments was used to identify the time delay between electron energy-loss events and photon emission in different materials: hBN flakes with 4.1 eV emitters, Au/SiO$_2$ core-shell nanoparticles and nanodiamonds with NV$^0$ centres \cite{Varkentina2022, Varkentina2023}[i, ii]. From the time-delay/energy-loss histograms of nanodiamonds (Fig. \ref{fig::CLE}B and F) and hBN (Fig. \ref{fig::CLE}C and E) the decay time of excitations as function of energy is measured. For the NV$^0$, the measured decay time (20-40 ns) matches the expected value from previous CL and PL experiments. For the 4.1 eV defect in hBN, the measured decay time is limited by the setup temporal response function ($\approx$ 2 ns). For both samples and Au/SiO$_2$ spheres (not shown), a fast decay trace is observed for energy losses in the visible range (Fig. \ref{fig::CLE}E and G labelled TR). This decay channel has been interpreted as transition radiation, which is emitted when a fast electron crosses an interface \cite{Abajo2010}. Decay times can also be measured in a setup which detects the arrival times of electrons without energy-loss selectivity, as shown for different scintillators in Ref. \cite{Yanagimoto2023}.

\begin{figure} [H]
\centering{\includegraphics[width=0.5\textwidth]{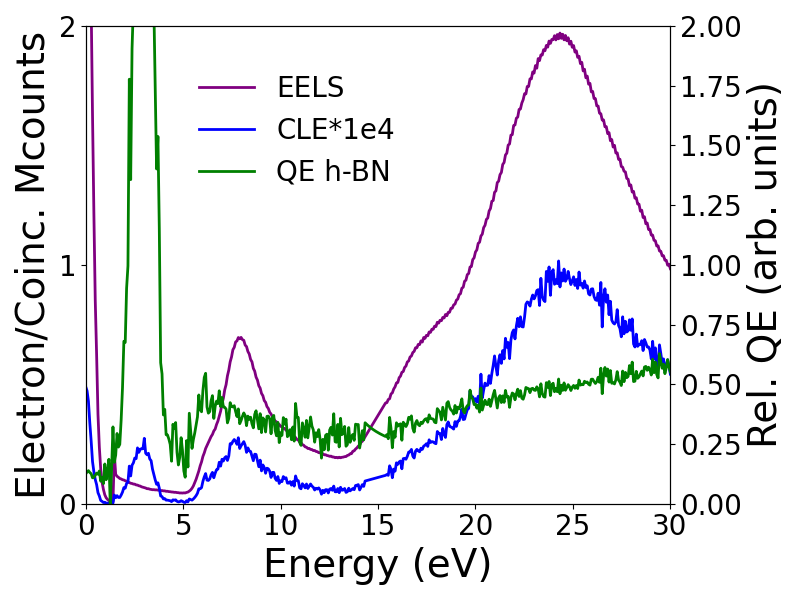}}
\caption{\textbf{EELS, CLE, relative quantum efficiency of hBN:} EELS (purple), CLE (blue), a sum of all electrons leading to a photon emission within the lifetime of the emitter analysed, and relative quantum efficiency (green) for an hBN flake containing 4.1 eV defects. Adapted with authorization from \cite{Varkentina2022}.
\label{fig::EELS-CLE-QE}}
\end{figure}

A time-integrated selection of electrons from electron-photon pairs with time delay below the typical decay time for a given excitation results in a spectrum which we name cathodoluminescence excitation (CLE) spectrum, in analogy to photoluminescence excitation spectroscopy (blue curve in Fig. \ref{fig::EELS-CLE-QE}), as it shows an emission intensity as a function of excitation energy.

From a CLE spectrum the most efficient energy losses towards photon emission can be identified. For example, for 4.1 eV defects in hBN bulk plasmon losses ($\approx$24 eV) are very efficient for photon generation (Fig. \ref{fig::EELS-CLE-QE}). The ratio of the CLE spectrum to the total EELS spectrum (purple curve in Fig. 9) is defined as the relative quantum efficiency (rQE) for light emission. From it (green curve in Fig. \ref{fig::EELS-CLE-QE}), the near-band edge of hBN (6.3 eV) is seen to be an efficient excitation pathway for 4.1 eV defect light emission. In most semiconductors tested with CLE, the rQE exhibits a higher value when higher energy is deposited in the material from the electron beam, reflecting the creation of a larger number of electron-hole pairs following these higher energy excitations.

Synchronized electron photon experiments are an effective way to improve visibility of signals, as it effectively removes the background from events which do not lead to photon emission \cite{Feist2022, Varkentina2022}. 

Electron-photon \cite{Preimesberger2025, Henke2025} experiments have demonstrated the entanglement between electrons and photons following inelastic scattering events. The heralded generation of non-classical light in electron scattering have been observed in electron-photon-photon \cite{Arend2024} synchronized measurements. A similar setup could be used to observe how the second order correlation function for the emitted light varies as a function of electron energy-loss. A variation is expected, as for small energy losses multiple electron-hole pairs cannot be created, a necessary condition for light bunching in CL experiments \cite{Meuret2015}. 

\subsection{Injection-EELS synchronization}
\label{sec::synchronized::injection-eels}

Besides the correlation of the emitted photon and the inelastically scattered electron, it is also possible to perform synchronized experiments between an injected laser photon field and EELS electrons. By doing so, one can perform PINEM while using standard continuous-gun electron microscopes. More specifically, ns-resolved laser sources offer an interesting choice because of its intense photonic field and narrowband spectral lines, while also being in the same range of temporal resolution given by the Timepix3 electron detector. This particular combination is suitable to perform EEGS spectroscopy \cite{Asenjo2013, Abajo2008b} as described above (see Section \ref{sec::comparative}).

\begin{figure} [H]
\centering{\includegraphics[width=0.5\textwidth]{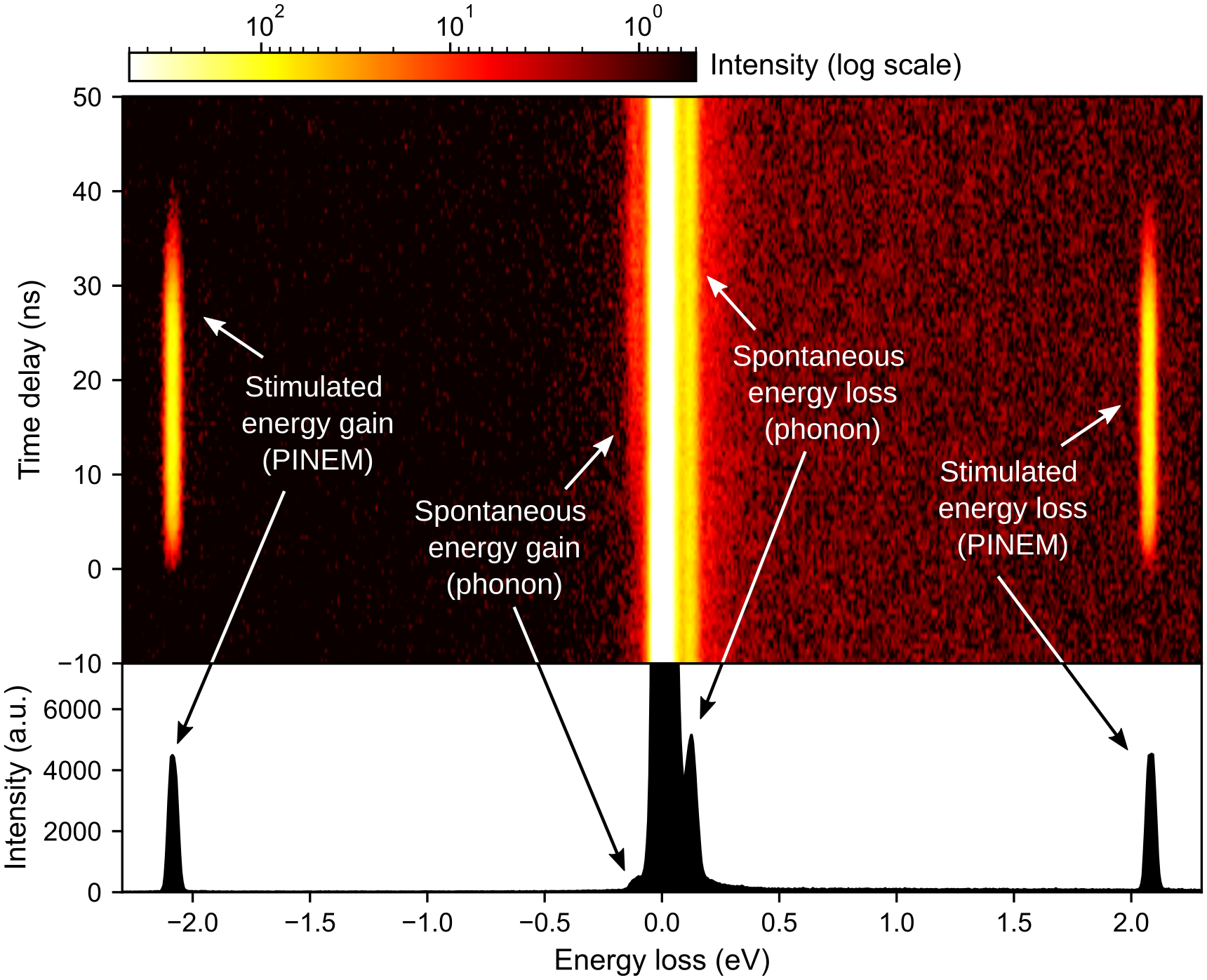}}
\caption{\textbf{PINEM with a TPX3:} Evolution of the EELS spectral signature of a thin SiN$_x$ window as a function of time after photoexcitation. The temporal delay is measured between the onset of the laser pulse and the arrival of the electrons, as recorded by the Timepix3 detector. The laser pulse generates both an evanescent optical field around the sample - enabling stimulated interaction with the electron beam and leading to PINEM signal - and heating of the material, which alters the gain/loss balance of the phonon-related intensity in the spectrum \cite{Castioni2025}.
\label{fig::PINEM}}
\end{figure}

Light-injection experiments in a continuous-gun electron microscope require a way to correlate the detected electrons with the injected laser pulse. Initial experiments have used an ns-electrostatic beam blanker placed in front of a ms-resolution EELS detector to do obtain a correlated signal with nanosecond resolution \cite{Das2019}. This experimental setup requires substantial modifications in the microscope spectrometer, but has the flexibility of being able to be performed with any kind of electron detector, regardless of its temporal resolution. Using event-based detectors, such as Timepix3, removes the need for the modified spectrometer, and the only requirement is to be able to timestamp laser pulses similarly as is done for electrons. An example of PINEM spectrum for a SiNx film, subject to a few tens of ns laser pulse and acquired on a Timepix3 detector, is shown in Fig. \ref{fig::PINEM}. The ns temporal resolution of the Timepix3 detector allows a fair assessment of the temporal shape of the laser pulse.

\subsection{Nanothermometry}
\label{sec::synchronized::nanothermometry}

PINEM experiments probe the optical excitations generated by a laser pulse. Following excitation, the excited carriers in the material can relax through non-radiative processes - most notably through interactions with phonons, resulting in energy dissipation and a rise in temperature. Therefore, beyond studying the optical properties, the material's out-of-equilibrium state can also be investigated dynamically by correlating the electrons scattered by the sample with the timing of the laser pulse.

In EELS spectroscopy, temperature changes can be monitored through various signatures across a broad spectral range, from the X-ray to the infrared. In metals, thermal expansion of the crystal lattice leads to a decrease in free electron density, which in turn causes a shift in the bulk plasmon energy. This method - commonly referred to as plasmon energy expansion thermometry (PEET) \cite{Mecklenburg2015, Chmielewski2020, Yang2025} - has been particularly explored in the context of in situ experiments, where accurately determining the material's temperature in such set up remains a challenge. In semiconductors, the bandgap energy tends to decrease with increasing temperature, which can be measured with sufficiently high energy resolution \cite{Tizei2016}. For any materials  the ratio between the energy gain and loss peaks associated with phonons is directly dependent to the sample's absolute temperature \cite{Lagos2018, Idrobo2018} (Fig. \ref{fig::PINEM}). It is also worth noting that diffraction techniques and emission spectroscopies, particularly cathodoluminescence, can be used to probe temperature evolution \cite{Mauser2021, Niekiel2017}.

\begin{figure} [H]
\centering{\includegraphics[width=0.5\textwidth]{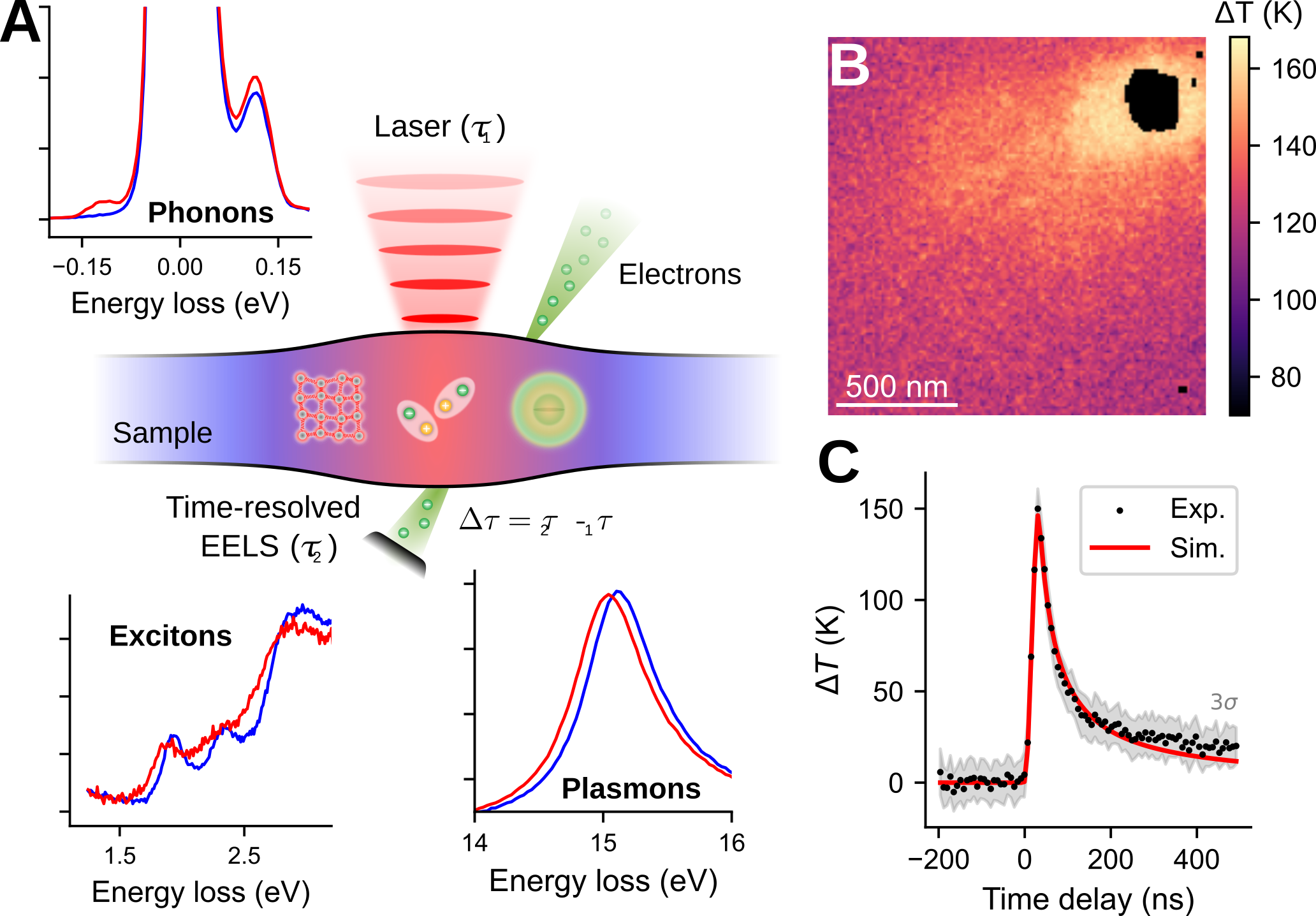}}
\caption{\textbf{Nanosecond-resolved nanothermometry:} Illustration of the photon-electron pump-probe experiment for nanosecond-resolved nanothermometry. (a) A 25 ns width pulsed laser in the visible range heats the sample, inducing absorption changes probed by EELS in the IR (phonons) to the soft X-ray (bulk plasmon) energy range in different material systems. (b) Temperature difference spatial mapping of an aluminium thin film following laser exposure. (c) Temperature temporal profile of the aluminium film depending on the time delay $\Delta t$ following the laser excitation, compared with a heat diffusion simulation in an infinite 2D film. Adapted with permission from \cite{Castioni2025}.
\label{fig::nanothermometry}}
\end{figure}

To date, most experiments investigating temperature variations in materials have relied on \textit{in situ} heating, particularly to induce localized temperature gradients over very short length scales \cite{Brintlinger2008, Shen2019, Park2024,Modrono2025}.  Some studies have demonstrated that the electron beam itself can serve as a highly localized heating source, with resulting temperature changes measured either via CL spectroscopy \cite{Mauser2021} or using thermocouples attached directly to the sample \cite{Kawamoto2018, Nguyen2024,Hettler2025}. While these static observations provide valuable insights, such approaches generally lack the temporal resolution needed to capture the dynamics of heat propagation at the nanometre scales, e.g. across interfaces.

The use of laser pulses to modify material temperature was proposed as early as the 1980s, with the primary goal of capturing phase transformations at the finest spatial resolution \cite{Hodgson1983, Takaoka1989}. Moreover, the use of short ($\approx$ps) laser pulses in pump-probe experiments has proven to be an effective method for generating coherent acoustic waves in nanostructures, resulting in lattice deformations that can be probed at the nanometer scale via diffraction \cite{Nakamura2020, Nakamura2022, Nakamura2023}. More recently, advances in instrumentation have enabled much greater control over laser injection, and have made it possible to accurately quantify temperature increases using diffraction and spectroscopy techniques \cite{Rossouw2013, Picher2015, Wu2018}and even through high-resolution imaging \cite{Uemura2022}. 

Time-resolved experiments using pulsed-source TEM have been conducted to investigate the out-of-equilibrium dynamics following photoexcitation of materials. By combining laser pulses in the fs to ns range with EELS, the evolution of atomic bonding and temperature in carbon-based materials have been tracked \cite{Carbone2009,Veen2015}. Similar experimental configurations have been used to study excitation dynamics in gold nanoparticles, where femtosecond laser pulses selectively excite interband transitions in the metal and resolved the different deexcitation pathways \cite{Kuwahara2022}. However, the extremely low probe current in these pulsed-source microscopes significantly limits the ability to spatially resolve spectroscopic signatures.

As an alternative, a continuous electron beam combined with synchronized electron detection has been shown to be highly effective for tracking temperature rise following laser heating \cite{Castioni2025}.  The laser is injected through a parabolic mirror, that focuses the beam onto a micrometre-sized spot. Depending on the material under investigation, laser-induced heating results in changes to the EELS spectral signature (Fig. \ref{fig::nanothermometry}). Thanks to the use of a focused electron probe and a temporal correlation system, it is possible to dynamically map temperature evolution with nanometre spatial resolution and nanosecond temporal resolution.

Fig. \ref{fig::nanothermometry}B shows an example of a temperature map of a thin aluminium film locally heated by the laser, where the temperature gradient clearly reflects the profile of the focused laser beam. Fig. \ref{fig::nanothermometry}C presents the temporal evolution of the film’s heating: the temperature rises by approximately 150 K above equilibrium shortly after the end of the laser pulse ($\approx$30 ns), then gradually decreases over several hundred nanoseconds. In this case, where the sample has a simple geometry, an analytical approach based on Fourier’s heat equations can be used to model the thermal behaviour. 

\subsection{Electron-electron coincidence experiments}
\label{sec::synchronized::electron-electron}

Although it has remained a niche topic, several studies investigating the correlation between EELS and SE - i.e., electrons originating from the sample, generated through interaction with the electron probe - were conducted from the late 1970s to the mid-1990s. The primary aim of this approach was to gain a better understanding of the physical origin of the signal resulting from fast electron/material interactions \cite{Liu1988}. By detecting these two signals in coincidence, it becomes possible to link specific absorption mechanisms (EELS) to the emission of SE.

To achieve this, SE are extracted and their energy measured either using a 180° hemispherical analyser or by converting their time-of-flight into an energy measurement. It is worth noting that extracting such low-energy electrons (typically $<$100 eV), emitted near the sample surface, is particularly challenging due to the strong magnetic field generated by the objective lens, which significantly distorts their trajectories \cite{Pijper1989,Kruit1988}. In the absence of time-resolved detectors, the arrival time of electrons can be determined either via a photomultiplier measuring the response of a scintillator or using MCP positioned behind the spectrometers.

In 1976, Voreades. reported the first coincidence experiments performed on a thin carbon film \cite{Vorades1976}. The effect of the material's work function ($\approx$eV) on the possibility of detecting SEs - since they must have enough energy to escape the surface—was clearly identified. Although the energy of the SE was not measured in this study, the paper highlighted that bulk plasmon decay was the primary source of SE production in these conductive thin films.

Fifteen years later, in 1991, Pijper and Kruit \cite{Pijper199} presented more advanced results on the same material, this time measuring the energy of SE. The authors demonstrated the presence of cascade excitation mechanisms, showing that SE production at a given energy increases with the deposited energy, with bulk and surface plasmon excitation as the dominant processes. More interestingly, they showed that this approach allows direct observation of ionized electrons from the material's valence band.

Subsequent studies explored excitation mechanisms in semiconductors. Scheinfein et al. \cite{Scheinfein1983} showed that, in the case of silicon, SE production is not driven by bulk plasmon decay but rather by excitation of valence bands located further below the Fermi level. Recent High Resolution Reflection EELS (HREELS) in coincidence with SE time-of-flight spectroscopy demonstrated that resonances above the vacuum level act as doorway states for secondary electron emission \cite{Niggas2025}. 
The issue of SE production following core-loss excitation has also been investigated. Complementing Voreades’s early results, which did not observe correlations at such high energy losses, Mullejans et al. \cite{Mullejans1992,Mullejans1993} demonstrated that carbon core-loss edges do not lead to increased SE production, regardless of energy. The same authors also showed that excitation in aloof mode can efficiently generate SE.

\section{Concluding remarks}

The purpose of this review was to describe recent temporal coincidence or synchronized electron spectroscopy experiments in the nanosecond to femtosecond range. A renewed interest in these techniques is occurring due to the development of advanced pulsed electron sources, precise light injection/collection systems for SEMs/STEMs, modern electron monochromators and time-resolved electron detectors. Their combination has significantly improved the quality and type of experiments possible in SEM/STEM for nano-optics.

The time-resolved event-based detector experiments described here can readily be extended beyond spectroscopy to include imaging and diffraction techniques, such as 4D-STEM or ptychography allowing for time-resolved experiment in SEMs/STEMs with continuous electron sources. Similarly, other stimuli instead of pulsed laser irradiation can be used, allowing for a large variety of time-resolved in situ experiments in similar microscopes.

The temporal resolution of the experiments with event-based detectors can be improved if better detectors are developed or if these are coupled with other technologies, such as fast blankers. For 60-300 keV electrons, the next generation of event-based electron detectors (Timepix4) will probably have a similar temporal resolution as the current generation (Timepix3), as the limitation is not the ASIC electronics but the jitter arising from electron drift in the detection layer \cite{Auad2024}.

The time-resolved experiments described here are still only possible in specialized laboratories, due to the technical difficulties in implementing time-resolved electron sources and detectors. A simplification and standardization of these methods would benefit the whole community and surely increase variety of experiments performed (new materials systems, new stimuli, for example).

\section{Funding}
This work was supported by the European Union’s Horizon programme through the project IMPRESS (grant agreement No. 101094299) and by the French National Agency for Research under the program of future investment TEMPOS-CHROMATEM (Reference ANR-10-EQPX-50) and the JCJC Grant SpinE (ANR-20-CE42-0020) and TPX4(ANR-23-CE42-0008).

\section{Data availability statement}
There is no original data associated to this review.

\section{Conflict of interest}
MK has licensed intellectual property to Attolight, which is producing the light injection/detection system used in some of the studies reviewed in this paper. The other authors declare that they have no conflict of interest.

\bibliography{Review_EELS-CL_Microscopy_bib}

@string{ACSNano={ACS\ Nano}}

@string{ACSPhot ={ACS\ Photon.}}

@string{AP    ={Adv.\ Phys.}}

@string{APL   ={Appl.\ Phys.\ Lett.}}

@string{CSR   ={Chem. Soc. Rev.}}

@string{JPCC  ={J.\ Phys.\ C}}

@string{JPChL ={J.\ Phys.\ Chem.\ Lett.}}

@string{NC    ={Nat.\ Commun.}}

@string{NJP   ={New\ J.\ Phys.}}

@string{NL    ={Nano\ Lett.}}

@string{NPhy  ={Nat.\ Phys.}}

@string{OL    ={Opt.\ Lett.}}

@string{PRB   ={Phys.\ Rev.\ B}}

@string{PRL   ={Phys.\ Rev.\ Lett.}}

@string{RMP   ={Rev.\ Mod.\ Phys.}}

@string{SS    ={Surf.\ Sci.}}

@article{Abajo2008,
  title = {Probing the Photonic Local Density of States with Electron Energy Loss Spectroscopy},
  author = {Garc\'{\i}a de Abajo, F. J. and Kociak, M.},
  journal = {Phys. Rev. Lett.},
  volume = {100},
  issue = {10},
  pages = {106804},
  numpages = {4},
  year = {2008},
  month = {Mar},
  publisher = {American Physical Society},
  doi = {10.1103/PhysRevLett.100.106804},
  url = {https://link.aps.org/doi/10.1103/PhysRevLett.100.106804}
}

@article{Abajo2008b,
doi = {10.1088/1367-2630/10/7/073035},
url = {https://dx.doi.org/10.1088/1367-2630/10/7/073035},
year = {2008},
month = {jul},
publisher = {},
volume = {10},
number = {7},
pages = {073035},
author = {García de Abajo, F J and Kociak, M},
title = {Electron energy-gain spectroscopy},
journal = {New Journal of Physics}
}

@article{Abajo2010,
  title={Optical excitations in electron microscopy},
  author={Garc{\'\i}a de Abajo, Francisco Javier},
  journal=RMP,
  url = {https://doi.org/10.1103/RevModPhys.82.209},
  volume={82},
  number={1},
  pages={209--275},
  year={2010},
  publisher={APS}
}

@article{Abajo2025,
  title={Roadmap for Quantum Nanophotonics with Free Electrons},
  author={de Abajo, F and Polman, Albert and Velasco, Cruz I and Kociak, Mathieu and Tizei, Luiz HG and St{\'e}phan, Odile and Meuret, Sophie and Sannomiya, Takumi and Akiba, Keiichirou and Auad, Yves and others},
  journal=ACSPhot,
  url = {https://doi.org/10.1021/acsphotonics.5c00585},
  year={2025}
}

@inproceedings{Ahn1984,
  title={Excited State Lifetime Measurement By EEL-CL Coincidence},
  author={Ahn, CC and Krivanek, OL},
  booktitle={Proceedings, annual meeting, Electron Microscopy Society of America},
  volume={43},
  pages={406--407},
  year={1984},
  organization={Cambridge University Press (CUP)}
}

@article{Arbouet2018,
  title={Ultrafast transmission electron microscopy: historical development, instrumentation, and applications},
  author={Arbouet, Arnaud and Caruso, Giuseppe M and Houdellier, Florent},
  journal={Advances in imaging and electron physics},
  url = {https://doi.org/10.1016/bs.aiep.2018.06.001},
  volume={207},
  pages={1--72},
  year={2018},
  publisher={Elsevier}
}

@misc{Arend2024,
      title={Electrons herald non-classical light}, 
      author={Germaine Arend and Guanhao Huang and Armin Feist and Yujia Yang and Jan-Wilke Henke and Zheru Qiu and Hao Jeng and Arslan Sajid Raja and Rudolf Haindl and Rui Ning Wang and Tobias J. Kippenberg and Claus Ropers},
      year={2024},
      eprint={2409.11300},
      archivePrefix={arXiv},
      primaryClass={quant-ph},
      url={https://arxiv.org/abs/2409.11300}, 
}

@article{Asenjo2013,
  title={Plasmon electron energy-gain spectroscopy},
  author={Asenjo-Garcia, Ana and De Abajo, FJ Garc{\'\i}a},
  journal=NJP,
  url = {https://doi.org/10.1088/1367-2630/15/10/103021},
  volume={15},
  number={10},
  pages={103021},
  year={2013},
  publisher={IOP Publishing}
}

@article{Auad2021,
  title={Unveiling the coupling of single metallic nanoparticles to whispering-gallery microcavities},
  author={Auad, Yves and Hamon, Cyrille and Tenc{\'e}, Marcel and Louren{\c{c}}o-Martins, Hugo and Mkhitaryan, Vahagn and St{\'e}phan, Odile and Garcia de Abajo, F Javier and Tizei, Luiz HG and Kociak, Mathieu},
  url = {https://doi.org/10.1021/acs.nanolett.1c03826},
  journal=NL,
  volume={22},
  number={1},
  pages={319--327},
  year={2021},
  publisher={ACS Publications}
}

@article{Auad2022,
  title={Event-based hyperspectral EELS: towards nanosecond temporal resolution},
  author={Auad, Yves and Walls, Michael and Blazit, Jean-Denis and St{\'e}phan, Odile and Tizei, Luiz HG and Kociak, Mathieu and de La Pe{\~n}a, Francisco and Tenc{\'e}, Marcel},
  journal={Ultramicroscopy},
  url = {https://doi.org/10.1016/j.ultramic.2022.113539},
  volume={239},
  pages={113539},
  year={2022},
  publisher={Elsevier}
}

@article{Auad2023,
  title={$\mu$eV electron spectromicroscopy using free-space light},
  author={Auad, Yves and Dias, Eduardo JC and Tenc{\'e}, Marcel and Blazit, Jean-Denis and Li, Xiaoyan and Zagonel, Luiz Fernando and St{\'e}phan, Odile and Tizei, Luiz HG and Garc{\'i}a de Abajo, F Javier and Kociak, Mathieu},
  url = {https://doi.org/10.1038/s41467-023-39979-0},
  journal=NC,
  volume={14},
  number={1},
  pages={4442},
  year={2023},
  publisher={Nature Publishing Group UK London}
}

@article{Auad2024,
  title={Time calibration studies for the Timepix3 hybrid pixel detector in electron microscopy},
  author={Auad, Yves and Baaboura, Jassem and Blazit, Jean-Denis and Tenc{\'e}, Marcel and St{\'e}phan, Odile and Kociak, Mathieu and Tizei, Luiz HG},
  journal={Ultramicroscopy},
  url = {https://doi.org/10.1016/j.ultramic.2023.113889},
  volume={257},
  pages={113889},
  year={2024},
  publisher={Elsevier}
}

@book{Ayache2010,
  title={Sample preparation handbook for transmission electron microscopy: techniques},
  author={Ayache, Jeanne and Beaunier, Luc and Boumendil, Jacqueline and Ehret, Gabrielle and Laub, Dani{\`e}le},
  url = {https://doi.org/10.1007/978-1-4419-5975-1},
  volume={2},
  year={2010},
  publisher={Springer Science \& Business Media}
}

@article{Bachu2024,
  title={Quantum Confined Luminescence in Two Dimensions},
  author={Bachu, Saiphaneendra and Habis, Fatimah and Huet, Benjamin and Woo, Steffi Y and Miao, Leixin and Reifsnyder Hickey, Danielle and Kim, Gwangwoo and Trainor, Nicholas and Watanabe, Kenji and Taniguchi, Takashi and others},
  journal=ACSPhot,
  url = {https://doi.org/10.1021/acsphotonics.4c01739},
  volume={12},
  number={1},
  pages={364--374},
  year={2024},
  publisher={ACS Publications}
}

@article{Barwick2009,
  title={Photon-induced near-field electron microscopy},
  author={Barwick, Brett and Flannigan, David J and Zewail, Ahmed H},
  url = {https://doi.org/10.1038/nature08662},
  journal={Nature},
  volume={462},
  number={7275},
  pages={902--906},
  year={2009},
  publisher={Nature Publishing Group UK London}
}

@article{Bezard2024,
  title={High-Efficiency coupling of free electrons to sub-$\lambda$3 modal volume, high-Q photonic cavities},
  author={B{\'e}zard, Malo and Si Hadj Mohand, Imene and Ruggierio, Luigi and Le Roux, Arthur and Auad, Yves and Baroux, Paul and Tizei, Luiz H G and Checoury, Xavier and Kociak, Mathieu},
  url = {https://doi.org/10.1021/acsnano.3c11211},
  journal=ACSNano,
  volume={18},
  number={15},
  pages={10417--10426},
  year={2024},
  publisher={ACS Publications}
}

@article{Boerrnert2023,
  title={A novel ground-potential monochromator design},
  author={Boerrnert, Felix and Uhlemann, Stephan and Mueller, Heiko and Gerheim, Volker and Haider, Maximilian},
  journal={Ultramicroscopy},
  url = {https://doi.org/10.1016/j.ultramic.2023.113805},
  volume={253},
  pages={113805},
  year={2023},
  publisher={Elsevier}
}

@article{Bonnet2021,
	title = {Nanoscale {Modification} of {WS2} {Trion} {Emission} by {Its} {Local} {Electromagnetic} {Environment}},
	volume = {21},
	issn = {1530-6984},
	url = {https://doi.org/10.1021/acs.nanolett.1c02600},
	doi = {10.1021/acs.nanolett.1c02600},
	number = {24},
	urldate = {2024-11-28},
	journal = {Nano Lett.},
	author = {Bonnet, No\'emie and Lee, Hae Yeon and Shao, Fuhui and Woo, Steffi Y. and Blazit, Jean-Denis and Watanabe, Kenji and Taniguchi, Takashi and Zobelli, Alberto and St\'ephan, Odile and Kociak, Mathieu and Gradečak, Silvija and Tizei, Luiz H. G.},
	month = dec,
	year = {2021},
	note = {Publisher: American Chemical Society},
	pages = {10178--10185},
}

@article{Borst1999,
  title={Time-autocorrelated two-photon counting technique for time-resolved fluorescence measurements},
  author={Borst, Walter L and Liu, Lin-I},
  journal={Rev. Sci. Instrum.},
  url = {https://doi.org/10.1063/1.1149540},
  volume={70},
  number={1},
  pages={41--49},
  year={1999},
  publisher={American Institute of Physics}
}

@article{Bourrellier2016,
  title={Bright UV single photon emission at point defects in h-BN},
  author={Bourrellier, Romain and Meuret, Sophie and Tararan, Anna and St{\'e}phan, Odile and Kociak, Mathieu and Tizei, Luiz HG and Zobelli, Alberto},
  journal=NL,
  url = {https://doi.org/10.1021/acs.nanolett.6b01368},
  volume={16},
  number={7},
  pages={4317--4321},
  year={2016},
  publisher={ACS Publications}
}

@article{Brintlinger2008,
  title={Electron thermal microscopy},
  author={Brintlinger, Todd and Qi, Yi and Baloch, Kamal H and Goldhaber-Gordon, David and Cumings, John},
  journal=NL,
  url = {https://doi.org/10.1021/nl0729375},
  volume={8},
  number={2},
  pages={582--585},
  year={2008},
  publisher={ACS Publications}
}

@article{Brown1956,
  title={Correlation between photons in two coherent beams of light},
  author={Brown, R Hanbury and Twiss, Richard Q},
  journal={Nature},
  url = {https://doi.org/10.1038/177027a0},
  volume={177},
  number={4497},
  pages={27--29},
  year={1956},
  publisher={Nature Publishing Group UK London}
}

@article{Castioni2025,
  title={Nanosecond nanothermometry in an electron microscope},
  author={Castioni, Florian and Auad, Yves and Blazit, Jean-Denis and Li, Xiaoyan and Woo, Steffi Y and Watanabe, Kenji and Taniguchi, Takashi and Ho, Ching-Hwa and St{\'e}phan, Odile and Kociak, Mathieu and others},
  journal=NL,
  url = {https://doi.org/10.1021/acs.nanolett.4c05692},
  volume={25},
  number={4},
  pages={1601--1608},
  year={2025},
  publisher={ACS Publications}
}

@article{
Carbone2009,
author = {Fabrizio Carbone  and Oh-Hoon Kwon  and Ahmed H. Zewail },
title = {Dynamics of Chemical Bonding Mapped by Energy-Resolved 4D Electron Microscopy},
journal = {Science},
volume = {325},
number = {5937},
pages = {181-184},
year = {2009},
doi = {10.1126/science.1175005},
URL = {https://www.science.org/doi/abs/10.1126/science.1175005}}

@article{Cha2010,
  title={Mapping local optical densities of states in silicon photonic structures with nanoscale electron spectroscopy},
  author={Cha, Judy J and Yu, Zongfu and Smith, Eric and Couillard, Martin and Fan, Shanhui and Muller, David A},
  url = {https://doi.org/10.1103/PhysRevB.81.113102},
  journal= PRB,
  volume={81},
  number={11},
  pages={113102},
  year={2010},
  publisher={APS}
}

@article{Chmielewski2020,
  title={Nanoscale temperature measurement during temperature controlled in situ TEM using Al plasmon nanothermometry},
  author={Chmielewski, A and Ricolleau, C and Alloyeau, D and Wang, Gang and Nelayah, J},
  journal={Ultramicroscopy},
  url = {https://doi.org/10.1016/j.ultramic.2019.112881},
  volume={209},
  pages={112881},
  year={2020},
  publisher={Elsevier}
}

@article{Das2019,
  title={Stimulated electron energy loss and gain in an electron microscope without a pulsed electron gun},
  author={Das, P and Blazit, JD and Tenc{\'e}, M and Zagonel, LF and Auad, Y and Lee, Yih Hong and Ling, Xing Yi and Losquin, A and Colliex, C and St{\'e}phan, O and others},
  journal={Ultramicroscopy},
  url = {https://doi.org/10.1016/j.ultramic.2018.12.011},
  volume={203},
  pages={44--51},
  year={2019},
  publisher={Elsevier}
}

@misc{Dellby2023,
  title={Ultra-high resolution {EELS} analysis and {STEM} imaging at 20 keV},
  author={Dellby, N and Quillin, SC and Krivanek, OL and Hrncirik, P and Mittelberger, A and Plotkin-Swing, B and Lovejoy, TC},
  url = {https://doi.org/10.1093/micmic/ozad067.305},
  year={2023},
  publisher={Oxford University Press US}
}

@article{Feist2015,
  title={Quantum coherent optical phase modulation in an ultrafast transmission electron microscope},
  author={Feist, Armin and Echternkamp, Katharina E and Schauss, Jakob and Yalunin, Sergey V and Sch{\"a}fer, Sascha and Ropers, Claus},
  url = {https://doi.org/10.1038/nature14463},
  journal={Nature},
  volume={521},
  number={7551},
  pages={200--203},
  year={2015},
  publisher={Nature Publishing Group UK London}
}

@article{Feist2022,
  title={Cavity-mediated electron-photon pairs},
  author={Feist, Armin and Huang, Guanhao and Arend, Germaine and Yang, Yujia and Henke, Jan-Wilke and Raja, Arslan Sajid and Kappert, F Jasmin and Wang, Rui Ning and Louren{\c{c}}o-Martins, Hugo and Qiu, Zheru and others},
  journal={Science},
  url = {https://doi.org/10.1126/science.abo5037},
  volume={377},
  number={6607},
  pages={777--780},
  year={2022},
  publisher={American Association for the Advancement of Science}
}

@article{Feldman2018,
  title = {Colossal photon bunching in quasiparticle-mediated nanodiamond cathodoluminescence},
  author = {Feldman, Matthew A. and Dumitrescu, Eugene F. and Bridges, Denzel and Chisholm, Matthew F. and Davidson, Roderick B. and Evans, Philip G. and Hachtel, Jordan A. and Hu, Anming and Pooser, Raphael C. and Haglund, Richard F. and Lawrie, Benjamin J.},
  journal = {Phys. Rev. B},
  volume = {97},
  issue = {8},
  pages = {081404},
  numpages = {5},
  year = {2018},
  month = {Feb},
  publisher = {American Physical Society},
  doi = {10.1103/PhysRevB.97.081404},
  url = {https://link.aps.org/doi/10.1103/PhysRevB.97.081404}
}

@article{Fiedler2023,
  title={Sub-to-super-Poissonian photon statistics in cathodoluminescence of color center ensembles in isolated diamond crystals},
  author={Fiedler, Saskia and Morozov, Sergii and Komisar, Danylo and Ekimov, Evgeny A and Kulikova, Liudmila F and Davydov, Valery A and Agafonov, Viatcheslav N and Kumar, Shailesh and Wolff, Christian and Bozhevolnyi, Sergey I and others},
  url = {https://doi.org/10.1515/nanoph-2023-0204},
  journal={Nanophotonics},
  volume={12},
  number={12},
  pages={2231--2237},
  year={2023},
  publisher={De Gruyter}
}

@article{Finot2021,
  title={Surface recombinations in III-nitride micro-LEDs probed by photon-correlation cathodoluminescence},
  author={Finot, Sylvain and Le Maoult, Corentin and Gheeraert, Etienne and Vaufrey, David and Jacopin, Gw{\'e}nol{\'e}},
  journal=ACSPhot,
  url = {https://doi.org/10.1021/acsphotonics.1c01339},
  volume={9},
  number={1},
  pages={173--178},
  year={2021},
  publisher={ACS Publications}
}

@book{Fox2010,
  title={Quantum Optics},
  author={Fox, Marx},
  url = {https://doi.org/10.1093/oso/9780198566724.003.0001},
  year={2006},
  publisher={Oxford Press}
}

@article{Gruber1997,
  title={Scanning confocal optical microscopy and magnetic resonance on single defect centers},
  author={Gruber, A and Drabenstedt, A and Tietz, C and Fleury, L and Wrachtrup, Joerg and Borczyskowski, C von},
  journal={Science},
  url = {https://doi.org/10.1126/science.276.5321.2012},
  volume={276},
  number={5321},
  pages={2012--2014},
  year={1997},
  publisher={American Association for the Advancement of Science}
}

@article{Guo2019,
  title={Radiation of dynamic toroidal moments},
  author={Guo, Surong and Talebi, Nahid and Campos, Alfredo and Kociak, Mathieu and van Aken, Peter A},
  journal=ACSPhot,
  url = {https://doi.org/10.1021/acsphotonics.8b01422},
  volume={6},
  number={2},
  pages={467--474},
  year={2019},
  publisher={ACS Publications}
}

@article{Guzzinati2017,
  title={Probing the symmetry of the potential of localized surface plasmon resonances with phase-shaped electron beams},
  author={Guzzinati, Giulio and B{\'e}ch{\'e}, Armand and Louren{\c{c}}o-Martins, Hugo and Martin, J{\'e}r{\^o}me and Kociak, Mathieu and Verbeeck, Jo},
  journal=NC,
  url = {https://doi.org/10.1038/ncomms14999},
  volume={8},
  number={1},
  pages={14999},
  year={2017},
  publisher={Nature Publishing Group UK London}
}

@article{Haak1984,
  title={Core-level electron-electron coincidence spectroscopy},
  author={Haak, HW and Sawatzky, GA and Ungier, L and Gimzewski, JK and Thomas, TD},
  journal={Rev. Sci. Instrum.},
  url = {https://doi.org/10.1063/1.1137823},
  volume={55},
  number={5},
  pages={696--711},
  year={1984}
}

@book{Hawkes2008,
  title={Science of microscopy},
  author={Hawkes, Peter W and Spence, John CH},
  year={2008},
  publisher={Springer Science \& Business Media}
}

@article{Henke2021,
  title={Integrated photonics enables continuous-beam electron phase modulation},
  author={Henke, Jan-Wilke and Raja, Arslan Sajid and Feist, Armin and Huang, Guanhao and Arend, Germaine and Yang, Yujia and Kappert, F Jasmin and Wang, Rui Ning and M{\"o}ller, Marcel and Pan, Jiahe and others},
  url = {https://doi.org/10.1038/s41586-021-04197-5},
  journal={Nature},
  volume={600},
  number={7890},
  pages={653--658},
  year={2021},
  publisher={Nature Publishing Group UK London}
}

@misc{Henke2025,
      title={Observation of quantum entanglement between free electrons and photons}, 
      author={Jan-Wilke Henke and Hao Jeng and Murat Sivis and Claus Ropers},
      year={2025},
      archivePrefix={arXiv},
      primaryClass={quant-ph},
      url={https://arxiv.org/abs/2504.13047}, 
}

@article{Hettler2025,
title = {Toward quantitative thermoelectric characterization of (nano)materials by in-situ transmission electron microscopy},
journal = {Ultramicroscopy},
volume = {268},
pages = {114071},
year = {2025},
issn = {0304-3991},
doi = {https://doi.org/10.1016/j.ultramic.2024.114071},
url = {https://www.sciencedirect.com/science/article/pii/S0304399124001505},
author = {Simon Hettler and Mohammad Furqan and Andres Sotelo and Raul Arenal}
}

@article{Hodgson1983,
    author = {Hodgson, R. T. and Boebinger, G. S. and Batson, P. E.},
    title = {In situ laser heating in a scanning transmission electron microscope},
    journal = {Appl. Phys. Lett.},
    volume = {43},
    number = {9},
    pages = {881-883},
    year = {1983},
    month = {11},
    issn = {0003-6951},
    doi = {10.1063/1.94505},
    url = {https://doi.org/10.1063/1.94505},
    
}

@article{Hou2021,
  title={Liquid-phase sintering of lead halide perovskites and metal-organic framework glasses},
  author={Hou, Jingwei and Chen, Peng and Shukla, Atul and Krajnc, Andra{\v{z}} and Wang, Tiesheng and Li, Xuemei and Doasa, Rana and Tizei, Luiz HG and Chan, Bun and Johnstone, Duncan N and others},
  url = {https://doi.org/10.1126/science.abf4460},
  journal={Science},
  volume={374},
  number={6567},
  pages={621--625},
  year={2021},
  publisher={American Association for the Advancement of Science}
}

@article{Husnik2013,
  title={Comparison of electron energy-loss and quantitative optical spectroscopy on individual optical gold antennas},
  author={Husnik, Martin and von Cube, Felix and Irsen, Stephan and Linden, Stefan and Niegemann, Jens and Busch, Kurt and Wegener, Martin},
  journal={Nanophotonics},
  url = {https://doi.org/10.1515/nanoph-2013-0031},
  volume={2},
  number={4},
  pages={241--245},
  year={2013},
  publisher={De Gruyter}
}

@article{Idrobo2018,
  title = {Temperature Measurement by a Nanoscale Electron Probe Using Energy Gain and Loss Spectroscopy},
  author = {Idrobo, Juan Carlos and Lupini, Andrew R. and Feng, Tianli and Unocic, Raymond R. and Walden, Franklin S. and Gardiner, Daniel S. and Lovejoy, Tracy C. and Dellby, Niklas and Pantelides, Sokrates T. and Krivanek, Ondrej L.},
  journal = {Phys. Rev. Lett.},
  volume = {120},
  issue = {9},
  pages = {095901},
  numpages = {4},
  year = {2018},
  month = {Mar},
  publisher = {American Physical Society},
  doi = {10.1103/PhysRevLett.120.095901},
  url = {https://link.aps.org/doi/10.1103/PhysRevLett.120.095901}
}

@article{Jannis2019,
  title={Spectroscopic coincidence experiments in transmission electron microscopy},
  author={Jannis, Daen and M{\"u}ller-Caspary, Knut and B{\'e}ch{\'e}, Armand and Oelsner, Andreas and Verbeeck, Johan},
  journal= APL,
  url = {https://doi.org/10.1063/1.5092945},
  volume={114},
  number={14},
  year={2019},
  publisher={AIP Publishing}
}

@article{Jannis2021,
  title={Coincidence detection of eels and edx spectral events in the electron microscope},
  author={Jannis, Daen and M{\"u}ller-Caspary, Knut and B{\'e}ch{\'e}, Armand and Verbeeck, Jo},
  journal={Appl. Sci.},
  url = {https://doi.org/10.3390/app11199058},
  volume={11},
  number={19},
  pages={9058},
  year={2021},
  publisher={MDPI}
}

@article{Kawamoto2018,
title = {Visualizing nanoscale heat pathways},
journal = {Nano Energy},
volume = {52},
pages = {323-328},
year = {2018},
issn = {2211-2855},
doi = {https://doi.org/10.1016/j.nanoen.2018.08.002},
url = {https://www.sciencedirect.com/science/article/pii/S2211285518305664},
author = {Naoyuki Kawamoto and Yohei Kakefuda and Isamu Yamada and Jianjun Yuan and Kotone Hasegawa and Koji Kimoto and Toru Hara and Masanori Mitome and Yoshio Bando and Takao Mori and Dmitri Golberg}
}

@article{Kawasaki2016,
  title={Extinction and scattering properties of high-order surface plasmon modes in silver nanoparticles probed by combined spatially resolved electron energy loss spectroscopy and cathodoluminescence},
  author={Kawasaki, Naohiko and Meuret, Sophie and Weil, Raphael and Louren{\c{c}}o-Martins, Hugo and St{\'e}phan, Odile and Kociak, Mathieu},
  url = {https://doi.org/10.1021/acsphotonics.6b00257},
  journal=ACSPhot,
  volume={3},
  number={9},
  pages={1654--1661},
  year={2016},
  publisher={ACS Publications}
}

@article{Kociak2014,
  title={Mapping plasmons at the nanometer scale in an electron microscope},
  author={Kociak, Mathieu and St{\'e}phan, Odile},
  journal=CSR,
  url = {https://doi.org/10.1039/C3CS60478K},
  volume={43},
  number={11},
  pages={3865--3883},
  year={2014},
  publisher={Royal Society of Chemistry}
}

@article{Kociak2017,
  title={Cathodoluminescence in the scanning transmission electron microscope},
  author={Kociak, Mathieu and Zagonel, Luis Fernando},
  journal={Ultramicroscopy},
  url = {https://doi.org/10.1016/j.ultramic.2017.03.014},
  volume={176},
  pages={112--131},
  year={2017},
  publisher={Elsevier}
}

@book{Kohl2008,
  title={Transmission electron microscopy},
  author={Kohl, Helmut and Reimer, Ludwig},
  booktitle={Transmission electron microscopy: Physics of image formation},
  pages={},
  year={2008},
  publisher={Springer}
}

@article{Konevcna2020,
  title={Electron beam aberration correction using optical near fields},
  author={Kone{\v{c}}n{\'a}, Andrea and de Abajo, F Javier Garc{\'\i}a},
  journal=PRL,
  url = {https://doi.org/10.1103/PhysRevLett.125.030801},
  volume={125},
  number={3},
  pages={030801},
  year={2020},
  publisher={APS}
}

@article{Krehl2018,
  title={Spectral field mapping in plasmonic nanostructures with nanometer resolution},
  author={Krehl, Jonas and Guzzinati, Giulio and Schultz, Johannes and Potapov, Pavel and Pohl, Darius and Martin, J{\'e}r{\^o}me and Verbeeck, Johan and Fery, Andreas and B{\"u}chner, Bernd and Lubk, Axel},
  url = {https://doi.org/10.1038/s41467-018-06572-9},
  journal=NC,
  volume={9},
  number={1},
  pages={4207},
  year={2018},
  publisher={Nature Publishing Group UK London}
}

@article{Krivanek2014,
  title={Vibrational spectroscopy in the electron microscope},
  author={Krivanek, Ondrej L and Lovejoy, Tracy C and Dellby, Niklas and Aoki, Toshihiro and Carpenter, Ray W and Rez, Peter and Soignard, Emmanuel and Zhu, Jiangtao and Batson, Philip E and Lagos, Maureen J and others},
  journal={Nature},
  url = {https://doi.org/10.1038/nature13870},
  volume={514},
  number={7521},
  pages={209--212},
  year={2014},
  publisher={Nature Publishing Group UK London}
}

@article{Kruit1984,
  title={Detection of X-rays and electron energy loss events in time coincidence},
  author={Kruit, P and Shuman, H and Somlyo, AP},
  url = {https://doi.org/10.1016/0304-3991(84)90199-2},
  journal={Ultramicroscopy},
  volume={13},
  number={3},
  pages={205--213},
  year={1984},
  publisher={Elsevier}
}

@article{Kruit1988,
title = {High-spatial-resolution surface-sensitive electron spectroscopy using a magnetic parallelizer},
journal = {Ultramicroscopy},
volume = {25},
number = {3},
pages = {183-193},
year = {1988},
issn = {0304-3991},
doi = {https://doi.org/10.1016/0304-3991(88)90013-7},
url = {https://www.sciencedirect.com/science/article/pii/0304399188900137},
author = {P. Kruit and J.A. Venables}
}

@article{Kuwahara2022,
    author = {Kuwahara, Makoto and Mizuno, Lira and Yokoi, Rina and Morishita, Hideo and Ishida, Takafumi and Saitoh, Koh and Tanaka, Nobuo and Kuwahara, Shota and Agemura, Toshihide},
    title = {Transient electron energy-loss spectroscopy of optically stimulated gold nanoparticles using picosecond pulsed electron beam},
    journal = {Applied Physics Letters},
    volume = {121},
    number = {14},
    pages = {143503},
    year = {2022},
    month = {10},
    issn = {0003-6951},
    doi = {10.1063/5.0108266},
    url = {https://doi.org/10.1063/5.0108266},
}

@article{Lagos2018,
  title={Thermometry with subnanometer resolution in the electron microscope using the principle of detailed balancing},
  author={Lagos, Maureen J and Batson, Philip E},
  journal=NL,
  url={https://doi.org/10.1021/acs.nanolett.8b01791},
  volume={18},
  number={7},
  pages={4556--4563},
  year={2018},
  publisher={ACS Publications}
}

@article{Liebtrau2021,
  title={Spontaneous and stimulated electron--photon interactions in nanoscale plasmonic near fields},
  author={Liebtrau, Matthias and Sivis, Murat and Feist, Armin and Louren{\c{c}}o-Martins, Hugo and Pazos-P{\'e}rez, Nicolas and Alvarez-Puebla, Ramon A and de Abajo, F Javier Garc{\'\i}a and Polman, Albert and Ropers, Claus},
  journal={Light. Sci. Appl.},
  url = {https://doi.org/10.1038/s41377-021-00511-y},
  volume={10},
  number={1},
  pages={82},
  year={2021},
  publisher={Nature Publishing Group UK London}
}

@article{Liu1988,
  title={Contrast and resolution of secondary electron images in a scanning transmission electron microscope},
  author={Liu, J and Cowley, JM},
  journal={Scanning Microscopy},
  volume={2},
  number={4},
  pages={12},
  year={1988}
}

@article{Losquin2015,
  title={Unveiling nanometer scale extinction and scattering phenomena through combined electron energy loss spectroscopy and cathodoluminescence measurements},
  author={Losquin, Arthur and Zagonel, Luiz F and Myroshnychenko, Viktor and Rodr{\'\i}guez-Gonz{\'a}lez, Benito and Tenc{\'e}, Marcel and Scarabelli, Leonardo and F\"orstner, Jens and Liz-Marz{\'a}n, Luis M and Garc{\'\i}a de Abajo, F Javier and St{\'e}phan, Odile and others},
  url={https://doi.org/10.1021/nl5043775},
  journal=NL,
  volume={15},
  number={2},
  pages={1229--1237},
  year={2015},
  publisher={ACS Publications}
}

@article{Losquin2015b,
  title={Link between cathodoluminescence and electron energy loss spectroscopy and the radiative and full electromagnetic local density of states},
  author={Losquin, Arthur and Kociak, Mathieu},
  url = {https://doi.org/10.1021/acsphotonics.5b00416},
  journal=ACSPhot,
  volume={2},
  number={11},
  pages={1619--1627},
  year={2015},
  publisher={ACS Publications}
}

@article{Lourencco2021,
  title={Optical polarization analogue in free electron beams},
  author={Louren{\c{c}}o-Martins, Hugo and G{\'e}rard, Davy and Kociak, Mathieu},
  url = {https://doi.org/10.1038/s41567-021-01163-w},
  journal=NPhy,
  volume={17},
  number={5},
  pages={598--603},
  year={2021},
  publisher={Nature Publishing Group UK London}
}

@article{Mahfoud2013,
  title={Cathodoluminescence in a scanning transmission electron microscope: A nanometer-scale counterpart of photoluminescence for the study of ii--vi quantum dots},
  author={Mahfoud, Zackaria and Dijksman, Arjen T and Javaux, Cl{\'e}mentine and Bassoul, Pierre and Baudrion, Anne-Laure and Plain, J{\'e}r{\^o}me and Dubertret, Beno{\^\i}t and Kociak, Mathieu},
  url = {https://doi.org/10.1021/jz402233x},
  journal= JPChL,
  volume={4},
  number={23},
  pages={4090--4094},
  year={2013},
  publisher={ACS Publications}
}

@article{Mauser2021,
  title={Employing cathodoluminescence for nanothermometry and thermal transport measurements in semiconductor nanowires},
  author={Mauser, Kelly W and Sol{\`a}-Garcia, Magdalena and Liebtrau, Matthias and Damilano, Benjamin and Coulon, Pierre-Marie and V{\'e}zian, St{\'e}phane and Shields, Philip A and Meuret, Sophie and Polman, Albert},
  url = {https://doi.org/10.1021/acsnano.1c00850},
  journal={ACS Nano},
  volume={15},
  number={7},
  pages={11385--11395},
  year={2021},
  publisher={ACS Publications}
}

@article{Mecklenburg2015,
  title={Nanoscale temperature mapping in operating microelectronic devices},
  author={Mecklenburg, Matthew and Hubbard, William A and White, ER and Dhall, Rohan and Cronin, Stephen B and Aloni, Shaul and Regan, BC},
  url = {https://doi.org/10.1126/science.aaa2433},
  journal={Science},
  volume={347},
  number={6222},
  pages={629--632},
  year={2015},
  publisher={American Association for the Advancement of Science}
}

@article{Meuret2015,
  title={Photon bunching in cathodoluminescence},
  author={Meuret, Sophie and Tizei, Luiz HG and Cazimajou, Thibauld and Bourrellier, Romain and Chang, Huan-Cheng and Treussart, F and Kociak, M},
  url = {https://doi.org/10.1103/PhysRevLett.114.197401},
  journal=PRL,
  volume={114},
  number={19},
  pages={197401},
  year={2015},
  publisher={APS}
}

@article{Meuret2016,
  title={Lifetime measurements well below the optical diffraction limit},
  author={Meuret, Sophie and Tizei, Luiz HG and Auzelle, Thomas and Songmuang, Rudee and Daudin, Bruno and Gayral, Bruno and Kociak, Mathieu},
  journal=ACSPhot,
  url = {https://doi.org/10.1021/acsphotonics.6b00212},
  volume={3},
  number={7},
  pages={1157--1163},
  year={2016},
  publisher={ACS Publications}
}

@article{Modrono2025,
author = {S\'aenz de Santa Mar\'ia Modro\~no, Pablo and Girard, Hugues A. and Arnault, Jean-Charles and Jacopin, Gw\'enol\'e},
title = {High-Resolution Thermal Sensing Using Temperature-Sensitive Cathodoluminescence Spectroscopy in Nitrogen-Doped Nanodiamonds},
journal = {physica status solidi (a)},
volume = {222},
number = {5},
pages = {2400573},
doi = {https://doi.org/10.1002/pssa.202400573},
url = {https://onlinelibrary.wiley.com/doi/abs/10.1002/pssa.202400573},
year = {2025}
}

@article{Mullejans1992,
  title = {Ratio between the energy-loss spectrum in coincidence with secondary electrons and the normal energy-loss spectrum for thin carbon films in the carbon K-edge region},
  author = {M\"ullejans, Harald and Bleloch, Andrew L.},
  journal = {Phys. Rev. B},
  volume = {46},
  issue = {13},
  pages = {8597--8599},
  numpages = {0},
  year = {1992},
  month = {Oct},
  publisher = {American Physical Society},
  doi = {10.1103/PhysRevB.46.8597},
  url = {https://link.aps.org/doi/10.1103/PhysRevB.46.8597}
}

@article{Mullejans1993,
title = {Secondary electron coincidence detection and time of flight spectroscopy},
journal = {Ultramicroscopy},
volume = {52},
number = {3},
pages = {360-368},
year = {1993},
issn = {0304-3991},
doi = {https://doi.org/10.1016/0304-3991(93)90047-2},
url = {https://www.sciencedirect.com/science/article/pii/0304399193900472},
author = {H. Mullejans and A.L. Bleloch and A. Howie and M. Tomita}
}

@article{Nakamura2020,
  title={Nanoscale imaging of unusual photoacoustic waves in thin flake VTe$_2$},
  author={Nakamura, Asuka and Shimojima, Takahiro and Chiashi, Yusuke and Kamitani, Manabu and Sakai, Hideaki and Ishiwata, Shintaro and Li, Han and Ishizaka, Kyoko},
  journal=NL,
  url = {https://doi.org/10.1021/acs.nanolett.0c01006},
  volume={20},
  number={7},
  pages={4932--4938},
  year={2020},
  publisher={ACS Publications}
}

@Article{Nakamura2022,
author ="Nakamura, A. and Shimojima, T. and Ishizaka, K.",
title  ="Visualizing optically-induced strains by five-dimensional ultrafast electron microscopy",
journal  ="Faraday Discuss.",
year  ="2022",
volume  ="237",
issue  ="0",
pages  ="27-39",
publisher  ="The Royal Society of Chemistry",
doi  ="10.1039/D2FD00062H",
url  ="http://dx.doi.org/10.1039/D2FD00062H"}

@article{Nakamura2023,
  title={Characterizing an optically induced sub-micrometer gigahertz acoustic wave in a silicon thin plate},
  author={Nakamura, Asuka and Shimojima, Takahiro and Ishizaka, Kyoko},
  journal=NL,
  url = {https://doi.org/10.1021/acs.nanolett.2c03938},
  volume={23},
  number={7},
  pages={2490--2495},
  year={2023},
  publisher={ACS Publications}
}

@article{Nelayah2007,
  title={Mapping surface plasmons on a single metallic nanoparticle},
  author={Nelayah, Jaysen and Kociak, Mathieu and St{\'e}phan, Odile and Garc{\'\i}a de Abajo, F Javier and Tenc{\'e}, Marcel and Henrard, Luc and Taverna, Dario and Pastoriza-Santos, Isabel and Liz-Marz{\'a}n, Luis M and Colliex, Christian},
  url = {https://doi.org/10.1038/nphys575},
  journal=NPhys,
  volume={3},
  number={5},
  pages={348--353},
  year={2007},
  publisher={Nature Publishing Group UK London}
}

@article{Nekula2025,
  title={Laser-based aberration corrector},
  author={Nekula, Zden{\v{e}}k and Juffmann, Thomas and Kone{\v{c}}n{\'a}, Andrea},
  url = {https://doi.org/10.48550/arXiv.2501.16501},
  journal={arXiv preprint arXiv:2501.16501},
  year={2025}
}

@article{Niekiel2017,
  title={Local temperature measurement in TEM by parallel beam electron diffraction},
  author={Niekiel, Florian and Kraschewski, Simon M and M{\"u}ller, Julian and Butz, Benjamin and Spiecker, Erdmann},
  journal={Ultramicroscopy},
  url = {https://doi.org/10.1016/j.ultramic.2016.11.028},
  volume={176},
  pages={161--169},
  year={2017},
  publisher={Elsevier}
}

@article{Niggas2025,
  title = {Identifying Electronic Doorway States in Secondary Electron Emission from Layered Materials},
  author = {Niggas, A. and Hao, M. and Richter, P. and Simperl, F. and Bl\"odorn, F. and Cap, M. and Kero, J. and Hofmann, D. and Bellissimo, A. and Burgd\"orfer, J. and Seyller, T. and Wilhelm, R. A. and Libisch, F. and Werner, W. S. M.},
  journal = {Phys. Rev. Lett.},
  volume = {135},
  issue = {16},
  pages = {166401},
  numpages = {8},
  year = {2025},
  month = {Oct},
  publisher = {American Physical Society},
  doi = {10.1103/qls7-tr4v},
  url = {https://link.aps.org/doi/10.1103/qls7-tr4v}
}

@book{Novotny2012,
  title={Principles of nano-optics},
  author={Novotny, Lukas and Hecht, Bert},
  url = {https://doi.org/10.1017/CBO9780511794193},
  year={2012},
  publisher={Cambridge university press}
}

@article{
Nguyen2024,
author = {Hieu Duy Nguyen  and Isamu Yamada  and Toshiyuki Nishimura  and Hong Pang  and Hyunyong Cho  and Dai-Ming Tang  and Jun Kikkawa  and Masanori Mitome  and Dmitri Golberg  and Koji Kimoto  and Takao Mori  and Naoyuki Kawamoto },
title = {STEM in situ thermal wave observations for investigating thermal diffusivity in nanoscale materials and devices},
journal = {Science Advances},
volume = {10},
number = {2},
pages = {eadj3825},
year = {2024},
doi = {10.1126/sciadv.adj3825},
URL = {https://www.science.org/doi/abs/10.1126/sciadv.adj3825}}

@article{Park2024,
  title={Nanoscale cathodoluminescence thermometry with a lanthanide-doped heavy-metal oxide in transmission electron microscopy},
  author={Park, Won-Woo and Olshin, Pavel K and Kim, Ye-Jin and Nho, Hak-Won and Mamonova, Daria V and Kolesnikov, Ilya E and Medvedev, Vassily A and Kwon, Oh-Hoon},
  url = {https://doi.org/10.1021/acsnano.3c10020},
  journal={ACS Nano},
  volume={18},
  number={6},
  pages={4911--4921},
  year={2024},
  publisher={ACS Publications}
}

@article{Pennycook1980,
  title={Observation of cathodoluminescence at single dislocations by STEM},
  author={Pennycook, SJ and Brown, LM and Craven, AJ},
  journal={Phil. Mag. A},
  url = {https://doi.org/10.1080/01418618008239335},
  volume={41},
  number={4},
  pages={589--600},
  year={1980},
  publisher={Taylor \& Francis}
}

@article{Picher2015,
title = {Vibrational and optical spectroscopies integrated with environmental transmission electron microscopy},
journal = {Ultramicroscopy},
volume = {150},
pages = {10-15},
year = {2015},
issn = {0304-3991},
doi = {https://doi.org/10.1016/j.ultramic.2014.11.023},
url = {https://www.sciencedirect.com/science/article/pii/S0304399114002393},
author = {Matthieu Picher and Stefano Mazzucco and Steve Blankenship and Renu Sharma}
}

@article{Pijper199,
  title={Detection of energy-selected secondary electrons in coincidence with energy-loss events in thin carbon foils},
  author={Pijper, Folbert J and Kruit, Pieter},
  journal=PRB,
  url = {https://doi.org/10.1103/PhysRevB.44.9192},
  volume={44},
  number={17},
  pages={9192},
  year={1991},
  publisher={APS}
}

@article{Pijper1989,
  title={Prospects for Photo Electron Spectroscopy in a Scanning Transmission Electron Microscope},
  author={Pijper, FJ and Bleeker, AJ and Endert, RJ and Kruit, P},
  journal={Scanning Microscopy},
  volume={3},
  number={1},
  pages={8},
  year={1989}
}

@misc{Preimesberger2025,
      title={Experimental Verification of Electron-Photon Entanglement}, 
      author={Alexander Preimesberger and Sergei Bogdanov and Isobel C. Bicket and Phila Rembold and Philipp Haslinger},
      year={2025},
      eprint={2504.13163},
      archivePrefix={arXiv},
      primaryClass={quant-ph},
      url={https://arxiv.org/abs/2504.13163}, 
}

@book{Schattschneider2012,
  title={Linear and chiral dichroism in the electron microscope},
  author={Schattschneider, Peter},
  url = {https://doi.org/10.1201/b11624},
  year={2012},
  publisher={CRC Press}
}

@article{Rossouw2013,
  title = {Structural and electronic distortions in individual carbon nanotubes under laser irradiation in the electron microscope},
  author = {Rossouw, David and Bugnet, Matthieu and Botton, Gianluigi A.},
  journal = PRB,
  volume = {87},
  issue = {12},
  pages = {125403},
  numpages = {5},
  year = {2013},
  month = {Mar},
  publisher = {American Physical Society},
  doi = {10.1103/PhysRevB.87.125403},
  url = {https://link.aps.org/doi/10.1103/PhysRevB.87.125403}
}

@article{Scheinfein1983,
  title = {Secondary-electron production pathways determined by coincidence electron spectroscopy},
  author = {Scheinfein, M. R. and Drucker, Jeff and Weiss, J. K.},
  journal = {Phys. Rev. B},
  volume = {47},
  issue = {7},
  pages = {4068--4071},
  numpages = {0},
  year = {1993},
  month = {Feb},
  publisher = {American Physical Society},
  doi = {10.1103/PhysRevB.47.4068},
  url = {https://link.aps.org/doi/10.1103/PhysRevB.47.4068}
}

@article{Schmidt2017,
  title={How dark are radial breathing modes in plasmonic nanodisks?},
  author={Schmidt, Franz-Philipp and Losquin, Arthur and Hofer, Ferdinand and Hohenau, Andreas and Krenn, Joachim R and Kociak, Mathieu},
  url = {https://doi.org/10.1021/acsphotonics.7b01060},
  journal=ACSPhot,
  volume={5},
  number={3},
  pages={861--866},
  year={2017},
  publisher={ACS Publications}
}

@article{Shao2022,
  title={Substrate influence on transition metal dichalcogenide monolayer exciton absorption linewidth broadening},
  author={Shao, Fuhui and Woo, Steffi Y and Wu, Nianjheng and Schneider, Robert and Mayne, Andrew J and De Vasconcellos, Steffen Michaelis and Arora, Ashish and Carey, Benjamin J and Preu{\ss}, Johann A and Bonnet, No{\'e}mie and others},
  journal={Phys. Rev. Mat.},
  url = {https://doi.org/10.1103/PhysRevMaterials.6.074005},
  volume={6},
  number={7},
  pages={074005},
  year={2022},
  publisher={APS}
}

@article{Shen2019,
    author = {Shen, Lang and Mecklenburg, Matthew and Dhall, Rohan and Regan, B. C. and Cronin, Stephen B.},
    title = {Measuring nanoscale thermal gradients in suspended MoS2 with STEM-EELS},
    journal = {Appl. Phys. Lett.},
    volume = {115},
    number = {15},
    pages = {153108},
    year = {2019},
    month = {10},
    issn = {0003-6951},
    doi = {10.1063/1.5094443},
    url = {https://doi.org/10.1063/1.5094443},
    
}

@article{Suenaga2010,
  title={Atom-by-atom spectroscopy at graphene edge},
  author={Suenaga, Kazu and Koshino, Masanori},
  journal={Nature},
  url = {https://doi.org/10.1038/nature09664},
  volume={468},
  number={7327},
  pages={1088--1090},
  year={2010},
  publisher={Nature Publishing Group UK London}
}

@article{Taleb2022,
  title={Charting the exciton--polariton landscape of WSe2 thin flakes by cathodoluminescence spectroscopy},
  author={Taleb, Masoud and Davoodi, Fatemeh and Diekmann, Florian K and Rossnagel, Kai and Talebi, Nahid},
  journal={Advanced Photonics Research},
  url = {https://doi.org/10.1002/adpr.202100124},
  volume={3},
  number={1},
  pages={2100124},
  year={2022},
  publisher={Wiley Online Library}
}

@article{Takaoka1989,
    author = {Takaoka, Akio and Nakamura, Noboru and Ura, Katsumi and Nishi, Hisami and Hata, Takao},
    title = {Local Heating of Specimen with Laser Diode in TEM},
    journal = {Journal of Electron Microscopy},
    volume = {38},
    number = {2},
    pages = {95-100},
    year = {1989},
    month = {01},
    issn = {0022-0744},
    doi = {10.1093/oxfordjournals.jmicro.a050731},
    url = {https://doi.org/10.1093/oxfordjournals.jmicro.a050731},
}

@article{Thomas2013,
  title={Imaging of high-Q cavity optical modes by electron energy-loss microscopy},
  author={Le Thomas, N and Alexander, DTL and Cantoni, M and Sigle, W and Houdr{\'e}, R and H{\'e}bert, C},
  journal=PRB,
  url = {https://doi.org/10.1103/PhysRevB.87.155314},
  volume={87},
  number={15},
  pages={155314},
  year={2013},
  publisher={APS}
}

@article{Tizei2013,
  title={Spatially resolved quantum nano-optics of single photons using an electron microscope},
  author={Tizei, LHG and Kociak, M},
  journal=PRL,
  url = {https://doi.org/10.1103/PhysRevLett.110.153604},
  volume={110},
  number={15},
  pages={153604},
  year={2013},
  publisher={APS}
}

@article{Tizei2016,
    author = {Tizei, Luiz H. G. and Lin, Yung-Chang and Lu, Ang-Yu and Li, Lain-Jong and Suenaga, Kazu},
    title = {Electron energy loss spectroscopy of excitons in two-dimensional-semiconductors as a function of temperature},
    journal = {Applied Physics Letters},
    volume = {108},
    number = {16},
    pages = {163107},
    year = {2016},
    month = {04},
    issn = {0003-6951},
    doi = {10.1063/1.4947058},
    url = {https://doi.org/10.1063/1.4947058},
    
}

@article{Tripathi2017,
  title={Cleaning graphene: Comparing heat treatments in air and in vacuum},
  author={Tripathi, Mukesh and Mittelberger, Andreas and Mustonen, Kimmo and Mangler, Clemens and Kotakoski, Jani and Meyer, Jannik C and Susi, Toma},
  journal={Phys. Stat. Sol. (RRL)},
  url = {https://doi.org/10.1002/pssr.201700124},
  volume={11},
  number={8},
  pages={1700124},
  year={2017},
  publisher={Wiley Online Library}
}

@article{Uemura2022,
title = {High-power laser irradiation for high-temperature in situ transmission electron microscopy},
journal = {Micron},
volume = {157},
pages = {103244},
year = {2022},
issn = {0968-4328},
doi = {https://doi.org/10.1016/j.micron.2022.103244},
url = {https://www.sciencedirect.com/science/article/pii/S0968432822000403},
author = {Naoki Uemura and Tomoya Egoshi and Koichi Murakami and Tokushi Kizuka}
}

@article{Varkentina2022,
  title={Cathodoluminescence excitation spectroscopy: Nanoscale imaging of excitation pathways},
  author={Varkentina, Nadezda and Auad, Yves and Woo, Steffi Y and Zobelli, Alberto and Bocher, Laura and Blazit, Jean-Denis and Li, Xiaoyan and Tenc{\'e}, Marcel and Watanabe, Kenji and Taniguchi, Takashi and others},
  url = {https://doi.org/10.1126/sciadv.abq4947},
  journal={Sci. Adv.},
  volume={8},
  number={40},
  pages={eabq4947},
  year={2022},
  publisher={American Association for the Advancement of Science}
}

@article{Varkentina2023,
    author = {Varkentina, Nadezda and Auad, Yves and Woo, Steffi Y. and Castioni, Florian and Blazit, Jean-Denis and Tenc\'e, Marcel and Chang, Huan-Cheng and Chen, Jeson and Watanabe, Kenji and Taniguchi, Takashi and Kociak, Mathieu and Tizei, Luiz H. G.},
    title = {Excitation lifetime extracted from electron–photon (EELS-CL) nanosecond-scale temporal coincidences},
    journal = {Applied Physics Letters},
    volume = {123},
    number = {22},
    pages = {223502},
    year = {2023},
    month = {12},
    issn = {0003-6951},
    doi = {10.1063/5.0165473},
    url = {https://doi.org/10.1063/5.0165473},
   
}

@article{Veen2015,
    author = {van der Veen, Renske M. and Penfold, Thomas J. and Zewail, Ahmed H.},
    title = {Ultrafast core-loss spectroscopy in four-dimensional electron microscopy},
    journal = {Structural Dynamics},
    volume = {2},
    number = {2},
    pages = {024302},
    year = {2015},
    month = {04},
    issn = {2329-7778},
    doi = {10.1063/1.4916897},
    url = {https://doi.org/10.1063/1.4916897},
    
}

@article{Yanagimoto2023,
  title={Time-correlated electron and photon counting microscopy},
  author={Yanagimoto, Sotatsu and Yamamoto, Naoki and Yuge, Tatsuro and Saito, Hikaru and Akiba, Keiichirou and Sannomiya, Takumi},
  url = {https://doi.org/10.1038/s42005-023-01371-1},
  journal={Comm. Phys.},
  volume={6},
  number={1},
  pages={260},
  year={2023},
  publisher={Nature Publishing Group UK London}
}

@article{Vorades1976,
title = {Secondary electron emission from thin carbon films},
journal = {Surface Science},
volume = {60},
number = {2},
pages = {325-348},
year = {1976},
issn = {0039-6028},
doi = {https://doi.org/10.1016/0039-6028(76)90320-4},
url = {https://www.sciencedirect.com/science/article/pii/0039602876903204},
author = {Demetrios Voreades}
}

@article{Yamamoto1984,
  title={Cathodoluminescence and polarization studies from individual dislocations in diamond},
  author={Yamamoto, N and Spence, JCH and Fathy, D},
  journal={Phil. Mag. B},
  url = {https://doi.org/10.1080/13642818408227648},
  volume={49},
  number={6},
  pages={609--629},
  year={1984},
  publisher={Taylor \& Francis}
}

@article{Yang2025,
  title={Improving the accuracy of temperature measurement on TEM samples using plasmon energy expansion thermometry (PEET): Addressing sample thickness effects},
  author={Yang, Yi-Chieh and Serafini, Luca and Gauquelin, Nicolas and Verbeeck, Johan and Jinschek, Joerg R},
  journal={Ultramicroscopy},
  url = {https://doi.org/10.1016/j.ultramic.2025.114102},
  volume={270},
  pages={114102},
  year={2025},
  publisher={Elsevier}
}

@article{Wang2018,
  title={Colloquium: Excitons in atomically thin transition metal dichalcogenides},
  author={Wang, Gang and Chernikov, Alexey and Glazov, Mikhail M and Heinz, Tony F and Marie, Xavier and Amand, Thierry and Urbaszek, Bernhard},
  journal=RMP,
  url = {https://doi.org/10.1103/RevModPhys.90.021001},
  volume={90},
  number={2},
  pages={021001},
  year={2018},
  publisher={APS}
}

@article{Werner2008,
  title={In-situ spectroscopic measurements of laser ablation-induced splitting and agglomeration of metal nanoparticles in solution},
  author={Werner, Daniel and Hashimoto, Shuichi and Tomita, Takuro and Matsuo, Shigeki and Makita, Yoji},
  journal=JPCC,
  url = {https://doi.org/10.1021/jp804647a},
  volume={112},
  number={43},
  pages={16801--16808},
  year={2008},
  publisher={ACS Publications}
}

@incollection{Williams2009,
  title={The transmission electron microscope},
  author={Williams, David B and Carter, C Barry},
  booktitle={Transmission electron microscopy: a textbook for materials science},
  url = {https://doi.org/10.1007/978-0-387-76501-3},
  year={2009},
  publisher={Springer}
}

@article{Woo2024,
  title={Nano-optics of transition metal dichalcogenides and their van der Waals heterostructures with electron spectroscopies},
  author={Woo, Steffi Y and Tizei, Luiz HG},
  url = {https://doi.org/10.1088/2053-1583/ad97c8},
  journal={2D Materials},
  volume={12},
  number={1},
  pages={012001},
  year={2024},
  publisher={IOP Publishing}
}

@article{Wu2018,
    author = {Wu, Yueying and Liu, Chenze and Moore, Thomas M and Magel, Gregory A and Garfinkel, David A and Camden, Jon P and Stanford, Michael G and Duscher, Gerd and Rack, Philip D},
    title = {Exploring Photothermal Pathways via in Situ Laser Heating in the Transmission Electron Microscope: Recrystallization, Grain Growth, Phase Separation, and Dewetting in Ag0.5Ni0.5 Thin Films},
    journal = {Microscopy and Microanalysis},
    volume = {24},
    number = {6},
    pages = {647-656},
    year = {2018},
    month = {12},
    issn = {1431-9276},
    doi = {10.1017/S1431927618015465},
    url = {https://doi.org/10.1017/S1431927618015465},
   
}

@article{Yanagimoto2025,
  title={Unveiling the nature of cathodoluminescence from photon statistics},
  author={Yanagimoto, Sotatsu and Yamamoto, Naoki and Yuge, Tatsuro and Sannomiya, Takumi and Akiba, Keiichirou},
  journal={Comm. Phys.},
  url = {https://doi.org/10.1038/s42005-025-01978-6},
  volume={8},
  number={1},
  pages={56},
  year={2025},
  publisher={Nature Publishing Group UK London}
}

@article{Zheng2017,
  title={Giant enhancement of cathodoluminescence of monolayer transitional metal dichalcogenides semiconductors},
  author={Zheng, Shoujun and So, Jin-Kyu and Liu, Fucai and Liu, Zheng and Zheludev, Nikolay and Fan, Hong Jin},
  journal=NL,
  url = {https://doi.org/10.1021/acs.nanolett.7b03585},
  volume={17},
  number={10},
  pages={6475--6480},
  year={2017},
  publisher={ACS Publications}
}

\end{document}